\documentclass[12pt]{iopart}
\usepackage{amsfonts}
\usepackage{graphicx}
\begin{document}
\title{Some partial solutions of Mathisson-Papapetrou equations in a
Schwarzschild field}
\author{Roman  Plyatsko}
\address{ Pidstryhach Institute of Applied Problems in Mechanics and
Mathematics\\ Ukrainian National Academy of Sciences, 3-b Naukova
Str.,\\ Lviv, 79060, Ukraine}

\ead{plyatsko@lms.lviv.ua}

\begin{abstract}

The analytical and numerical solutions of the Mathisson-Papapetrou equations under the
Mathisson-Pirani supplementary condition describing highly relativistic
(ultrarelativistic) motions of a spinning particle in a Schwarzschild field are
investigated. The known condition $|S_0|/mr\ll 1$, which is necessary for a test
particle, holds on all these solutions. The explicit expressions for the non-equatorial
circular orbits, in particular for the space boundaries of the region of existence of
these orbits, are obtained. The dynamics of the deviation of a spinning particle from the
equatorial ultrarelativistic circular orbit with $r=3M$ caused by the non-zero initial
value of the radial particle's velocity is studied. It is shown in the concrete cases
that spin can considerable influence the shape of an ultrarelativistic trajectory, as
compare to the corresponding geodesic trajectory, for the short time, less than the time
of one or two revolutions of a spinning particle around a Schwarzschild mass.

\end{abstract}


\maketitle
\section {Introduction}

The present paper deals with the Mathisson-Papapetrou equations describing the motion of
a spinning test particle in a gravitational field\footnote[1]{It was pointed out in some
papers that these equations were first derived by A.Papapetrou. Perhaps, such a
conclusion was inspired by the non-adequate sentence from \cite{2}, p. 254. About the
role of Dr. Myron Mathisson (14.12.1897, Warsaw -- 13.09.1940, England; P.Dirac published
Obituary in {\it Nature} 1940, {\bf 146}, 613) and his dramatic life see \cite{3}. After
\cite{1} these equations were rederived and reformulated in more than 20 papers, many of
them are cited in \cite{19, 28}} \cite{1,2}. Partial solutions of these equations have
been considered since 1951 \cite{4}. Different aspects of the orbital motions of a
spinning test particle in the gravitational fields were investigated in many papers
[5-41]. For the correct description of the trajectories of a spinning test particle by
the Mathisson-Papapetrou equations it is necessary to take into account the condition
$|S_0|/Mm\ll 1$ ($|S_0|$ and $m$ are the absolute value of spin and the mass of a
particle respectively, and $M$ is the mass of a gravitational source). Probably, R.Wald
was the first who introduced this condition explicitly \cite{9}. Another point of
importance concerning the Mathisson-Papapetrou equations is the supplementary condition
for the concretization of the center mass position of a spinning particle. In relativity,
the position of the center of mass of a rotating body depends on the frame of reference
[42-44]. C.M\"oller distinguished the proper center of mass and the non-proper centers of
mass. Because the correct definition of the center of mass for a spinning particle is a
subject of discussion, different conditions are used. Most known are the Mathisson-Pirani
\cite{1,45} and the Tulczyjew-Dixon conditions \cite{46, 47} (the first of them is often
called as the Pirani condition, though before \cite{45} it was used in \cite{1}). More
rare is the Corinaldesi-Papapetrou condition \cite{4}. (All these conditions are
described in section 2). For example, the Mathisson-Pirani condition was used in [8, 19,
33, 40] and the Tulczyjew-Dixon condition was used in [9, 12, 24-26, 28-30, 32, 35, 36,
39, 41]. Authors of some papers take into account both the Mathisson-Pirani and
Tulczyjew-Dixon conditions [18, 20, 37, 38]. More about the supplementary conditions we
shall write below. Here we stress that it is important to study the solutions of the
Mathisson-Papapetrou equations under different conditions and to compare their physical
consequences.

 The feature of importance of the Mathisson-Papapetrou equations is that they
are obtained without the restriction on the particle's velocity. This point is common for
these equations and the geodesic equations. However, it is easy to see the essential
difference: the highly relativistic orbits of a spinless test particle following from the
geodesic equations have been studied in full detail in different gravitational fields
whereas the possible orbits of a fast particle following from the Mathisson-Papapetrou
equations are investigated insufficiently even in a Schwarzschild field. The reason for
this situation is connected with the {\it a priori} assumption that the world lines of a
spinning test particle are close to the corresponding geodesic lines in the whole range
of the particle's velocity. Such an assumption is acceptable for the low velocity of a
spinning particle but is not evident for the high velocity when {\it a priori} we cannot
exclude the specific influence of the great relativistic Lorentz $\gamma$- factor on the
interaction of the spin with a gravitational field.

It has been shown in \cite{27,33} that, from the point of view of
the observer comoving with a fast spinning test particle, the
3-acceleration of this particle relative to the corresponding
spinless particle (i.e. the particle with the same initial values
of the coordinates and velocity) in a Schwarzschild field is
proportional to $\gamma^2$. In this sense we say that the highly
relativistic motions (with $\gamma^2\gg 1$) of a spinning test
particle essentially differ from the geodesic motions. (Other
possible features of the "essentially non-geodesic motions" of a
spinning particle we shall consider below). The force deviating
the motion of a spinning particle from the geodesic motion is the
gravitational ultrarelativistic spin-orbit force \cite{27,33}.

Naturally, further investigations must be carried out to determine
the possible deviation of the orbital motion of a spinning
particle from the corresponding geodesic orbits (i.e. the orbits
with the same initial values of the coordinates and velocity) from
the point of view of the observer at rest relative to a
Schwarzschild source.

The purpose of this paper is to present the new partial solutions of the
Mathisson-Papapetrou equations under the Mathisson-Pirani supplementary condition in a
Schwarzschild field describing the highly relativistic motions of a spinning test
particle. We shall consider both the analytical and numerical solutions. In this context
we stress that the wide classes of the numerical solutions of the Mathisson-Papapetrou
equations under the Tulczyjew-Dixon condition in the Schwarzschild and Kerr fields were
investigated in [24-26, 30, 35, 39]. The results of these papers are of importance for
study of the dynamics of the binary black holes. We point out some differences between
these results and our approach: (1) we use the Mathisson-Pirani supplementary condition
and argue this choice; (2) many results of [24-26, 30, 35, 39] describe the situations
when the value $|S_0|/Mm$ is of order 1, i.e. when the Wald condition is not satisfied.
Whereas all our solutions satisfy the relationship $|S_0|/Mm\ll 1$; (3) we focus our
attention on the highly relativistic motion just of a microscopic spinning particle,
because in reality any macroscopic test body cannot be considered as highly relativistic.
A common feature of all our solutions is that they describe the ultrarelativistic motions
with the $\gamma$- factor satisfying the relationship: $\gamma^{-2}$ is of order
$|S_0|/Mm$.

This paper is organized as follows. In section 2 we review the
Mathisson-Papapetrou equations and the supplementary conditions.
The partial analytical solutions of these equations describing
non-equatorial circular orbits of a spinning test particle in a
Schwarzschild field are studied in section 3. In section 4, using
the integrals of the energy and angular momentum in a
Schwarzschild field, we write the strict Mathisson-Papapetrou
equations under the Mathisson-Pirani supplementary condition in
the form of the second-order differential equations relative to
the coordinates $r$ and $\varphi$ for the equatorial motions of a
spinning particle. The numerical solutions of these equations are
investigated in section 5.

Throughout this paper we use units $c=G=1$. Greek indices run
1,2,3,4 and latin indices 1,2,3; the signature of the metric
(--,--,--,+) is chosen.

\section{The Mathisson-Papapetrou equations and supplementary conditions}

The Mathisson-Papapetrou equations can be written in the form
\cite{1, 2}
\begin{equation}\label{1}
\frac D {ds} \left(mu^\lambda + u_\mu\frac {DS^{\lambda\mu}}
{ds}\right)= -\frac {1} {2} u^\pi S^{\rho\sigma}
R^{\lambda}_{\pi\rho\sigma},
\end{equation}
\begin{equation}\label{2}
\frac {DS^{\mu\nu}} {ds} + u^\mu u_\sigma \frac {DS^{\nu\sigma}}
{ds} - u^\nu u_\sigma \frac {DS^{\mu\sigma}} {ds} = 0
\end{equation}
where $u^\lambda$ is the 4-velocity of a spinning particle,
$S^{\mu\nu}$ is the antisymmetric tensor of spin, $m$ and $D/ds$
are, respectively, the mass and the covariant derivative with
respect to proper time $s$; $R^{\lambda}_{\pi\rho\sigma}$ is the
Riemann curvature tensor of the spacetime. The Mathisson-Pirani
supplementary condition for equations (\ref{1}), (\ref{2}) is
\cite{1,45}
\begin{equation}\label{3}
S^{\mu\nu} u_\nu = 0.
\end{equation}
(Relationship (\ref{3}) was used earlier in special-relativistic
electrodynamics \cite{48}).

The Tulczyjew-Dixon supplementary condition is \cite{46,47}
\begin{equation}\label{4}
S^{\mu\nu} P_\nu = 0
\end{equation}
where
\begin{equation}\label{5}
P^\nu = mu^\nu + u_\mu\frac {DS^{\nu\mu}}{ds}
\end{equation}
is the particle's 4-momentum.

Papapetrou and Corinaldesi used the condition \cite{4}
\begin{equation}\label{6}
S^{i4}  = 0.
\end{equation}
(It is easy to see that for $u_i=0$ relationship (\ref{6})
coincides with (\ref{3})).

It is known from \cite{49,50} that in the Minkowski spacetime the
Mathisson-Papapetrou equations under condition (\ref{3}) have, in
addition to the usual solutions describing the straight
worldlines, a family of solutions describing the helical
worldlines. (As a partial case, this family contain the circular
solutions). These unusual solutions often are called as the
Weyssenhoff orbits. By C.M\"oller, the usual solutions describe
the motion of the proper center of mass of a spinning body
(particle), and the helical solutions describe the motions of the
family of the non-proper centers of mass \cite{43,44}.

To avoid the superfluous solutions of equations (\ref{1}),
(\ref{2}), instead of condition (\ref{3}) the Tulczyjew-Dixon
condition was introduced. This condition picks out a unique
worldline. However, the question arises: is this worldline close,
in the certain sense, to the usual worldline of equations
(\ref{1}), (\ref{2}) under condition (\ref{3})? It is simple to
answer this question when the relationship
\begin{equation}\label{7}
m|u^\nu|\gg \left|u_\mu\frac {DS^{\nu\mu}}{ds}\right|
\end{equation}
takes place, because in this case conditions (\ref{4}) practically coincides with
(\ref{3}). It is easy to check that relationship (\ref{7}) holds if the Wald condition
\begin{equation}\label{8}
|\varepsilon|\ll 1, \quad \varepsilon\equiv \frac{S_0}{Mm}
\end{equation}
is taken into account and, in addition, if the particle's velocity
is not too close to the velocity of light\footnote[1]{For
different estimates it is helpful to write numbers of
$|\varepsilon|$ for various particles. It is known that in units
where $c=G=1$ the numerical values of the electron's mass, the
Sun's mass and the Planck constant are equal respectively to:
$m=2.3\cdot10^{-66}$, $M=5\cdot 10^{-6}$, $h=2\cdot 10^{-86}$
(see, e.g.\cite{50a}). Then for a Schwarzschild black hole of mass
that is equal to three of the Sun's mass we have
$|\varepsilon_e|=4.6\cdot 10^{-17}$. The analogous values for a
proton and a neutrino (of mass corresponding to the energy of
order $1eV$) are: $|\varepsilon_p|=2.5\cdot 10^{-20}$,
$|\varepsilon_\nu|=2.3\cdot 10^{-11}$. By these values it is easy
to recalculate the corresponding values for a massive
Schwarzschild black hole or for a hypothetic microscopic black
hole.}. Here we point out that there is the connection between
$S_0$ snd $S^{\mu\nu}$
\begin{equation}\label{9}
S_0^2=\frac12S_{\mu\nu}S^{\mu\nu}.
\end{equation}

Another situation with relationship (\ref{7}) takes place for the highly relativistic
motions of a spinning particle when $u^\nu$ is proportional to the $\gamma$-factor,
because the term $DS^{\nu\mu}/ds$ in (\ref{7}) depends on $u^\nu$ and on $\gamma$-factor
as well. Therefore, in general, for the ultrtelativistic particle's velocity it is not
obvious that (\ref{7}) is valid.

So, if the velocity of a spinning particle is not too high, one can "forget" about the
Mathisson-Pirani supplementary condition, because it is sufficient to use the
Tulczyjew-Dixon condition. Whereas for highly relativistic motions the Mathisson-Pirani
condition must be taken into account. We stress that just the Mathisson-Pirani condition
is derived in some papers by different method \cite{8,51,52}. (For example, we agree with
the statement that the Mathisson-Pirani condition "... arises in a natural fashion in the
course of the derivation", \cite{51}, p.112). That is, the Mathisson-Pirani condition is
the necessary one, though often, with high accuracy, it can be substituted by the
Tulczyjew-Dixon condition.

We summarize: the existence of the superfluous solutions of the
Mathisson-Papapetrou equations under the Mathisson-Pirani
supplementary condition is not a reason to ignore this condition.
Because just among all solutions of these equations under
Mathisson-Pirani condition there is the single solution describing
the motion of the particle's proper center of mass. As we shall
see in section 5, it is not a great problem to recognize  such a
solution, at least in some cases of importance.

In this paper we shall use the Mathisson-Pirani condition. At the
same time, in some points, we take into account condition
(\ref{4}). More exactly, for the concrete solutions of the
Mathisson-Papapetrou equations we shall estimate the value of the
left-hand side of relationship (\ref{4}).

\section{Non-equatorial circular orbits of a spinning particle
 in a Schwarzschild field}

The subject of this section is inspired by the discussion in
papers \cite{53, 54} where the question was considered concerning
the existence or non-existence of non-equatorial circular
geodesics with constant latitude in a Kerr metric. It was shown in
\cite{54} that such geodesics do not exist. An interesting
question is the following: are there not the non-equatorial
circular orbits with constant latitude in the Schwarzschild or
Kerr metric according to the Mathisson-Papapetrou equations? Now
we shall consider the case of a Schwarzschild metric.

Here and in the following the standard Schwarzschild coordinates $x^1=r,\quad
x^2=\theta,\quad x^3=\varphi,\quad x^4=t$ are used, and we put $r>2M$ ($M$ is the
Schwarzschild mass), i.e. the region above the horizon surface is under investigation.

Let us check, by the direct calculations, do equations (\ref{1}),
(\ref{2}) have the solutions with
\begin{equation}
\label{10} r=const\ne 0, \quad \theta=const\ne 0, \pi/2, \pi,
\end{equation}
\begin{equation}
\label{11} u^3=\frac{d\varphi}{ds}=const\ne 0, \quad
u^4=\frac{dt}{ds}=const\ne 0.
\end{equation}
Relationships (\ref{10}), (\ref{11}) correspond to the circular
motions with the constant particle's velocity.

We start from equations (\ref{2}).

\subsection{The consideration of equations (\ref{2})}

It is convenient to use the representation of equations (\ref{2})
through the 3-vector of spin $S_i$ where by definition \cite{40}
\begin{equation}\label{12}
 S_i=\frac{1}{2}\sqrt{-g}\varepsilon_{ikl} S^{kl}
\end{equation}
($\varepsilon_{ikl}$ is the spatial Levi-Civita symbol). The
inverse relationship is
\begin{equation}\label{13}
S^{kl}=\frac{1}{\sqrt{-g}}\varepsilon^{klm}S_m.
\end{equation}
By condition (\ref{3}) we have
\begin{equation}\label{14}
 S^{i4}=\frac{u_k}{u_4}S^{ki}.
\end{equation}
(We point out that in many papers the 4-vector of spin $s_\lambda$
is used where
\begin{equation}\label{15}
 s_\lambda=\frac{1}{2}\sqrt{-g}\varepsilon_{\lambda\mu\nu\sigma}u^\mu
 S^{\nu\sigma}.
\end{equation}
The relationship between $S_i$ and $s_\lambda$ is
$S_i=u_is_4-u_4s_i$ \cite{40}).

According to \cite{40} three
independent equations from (\ref{2}) may be written in the form
\begin{equation}\label{16}
u_4\dot S_i-\dot u_4 S_i+2\left(\dot u_{[4} u_{i]}-u^\pi u_\rho
\Gamma^\rho_{\pi [4} u_{i]}\right)S_k u^k+2S_n\Gamma^n_{\pi[4}
u_{i]} u^\pi =0.
\end{equation}
Here and in the following a dot denotes usual differentiation with
respect to proper time $s$, and square brackets denote
antisymmetrization of indices.  We stress that equations (\ref{2})
under condition (\ref{3}) is equivalent to equations (\ref{16}).

Taking into account equations (\ref{10}), (\ref{11}) we write three equations from
(\ref{16}) as
\begin{equation}
\label{17} \dot S_1+S_3u^3u^4u_4(\Gamma^4_{14}-\Gamma^3_{13})=0,
\end{equation}
\begin{equation}
\label{18} \dot S_2-S_3u^3u^4u_4\Gamma^3_{23}=0,
\end{equation}
\begin{equation}
\label{19} \dot
S_3+S_1u_3(g^{44}\Gamma^1_{44}-g^{33}\Gamma^1_{33})-
S_2\Gamma^2_{33}u_3g^{33}=0.
\end{equation}
It is easy to see that the set of equations (\ref{17})-(\ref{19})
has the partial solution with
\begin{equation}
\label{20} S_3\equiv S_{\varphi}=0, \quad S_1\equiv S_r=const,
\quad S_2\equiv S_\theta=const,
\end{equation}
\begin{equation}
\label{21}
S_1(g^{44}\Gamma^1_{44}-g^{33}\Gamma^1_{33})-S_2g^{33}\Gamma^2_{33}=0.
\end{equation}
Using the explicit expressions for $g^{\mu\nu}$ and
$\Gamma^{\pi}_{\rho\sigma}$ in the standard Schwarzschild
coordinates we rewrite equation (\ref{21})
\begin{equation}
\label{22} S_1\left(1-\frac{3M}{r}\right)+S_2\frac{\cos\theta}{r\sin\theta}=0.
\end{equation}
In the following we shall analyse equations (\ref{1})  for the case just when
relationships (\ref{20}), (\ref{22}) take place.

\subsection{The consideration of equations (\ref{1})}

Let us write equations (\ref{1}) taking into account relationships (\ref{10}),
(\ref{11}), (\ref{13}), (\ref{14}). Using the explicit expressions for
$R^{\lambda}_{\pi\rho\sigma},$  two equations of set (\ref{1}) with $\lambda=1$ and
$\lambda=2$ can be written as
$$
m(\Gamma^1_{33}u^3u^3+\Gamma^1_{44}u^4u^4)+u^3(\Gamma^1_{33}-
g^{44}g_{33}\Gamma^1_{44})\frac{1}{\sqrt{-g}}(g_{44}\Gamma^4_{14}u^4u^4S_2+
$$
\begin{equation}
\label{23} +g_{33}\Gamma^3_{13}u^3u^3S_2-g_{33}\Gamma^3_{23}u^3u^3S_1)=
-\frac{3M}{r^3}u^3S_2\sin\theta,
\end{equation}
$$
m\Gamma^2_{33}u^3u^3+ u^3\Gamma^2_{33}\frac{1}{\sqrt{-g}}(g_{44}\Gamma^4_{14}u^4u^4S_2+
+g_{33}\Gamma^3_{13}u^3u^3S_2-
$$
\begin{equation}
\label{24} -g_{33}\Gamma^3_{23}u^3u^3S_1)=-\frac{3M}{r^3}u^3S_1\sin\theta.
\end{equation}
Two other equations of set (\ref{1}), with $\lambda=3$ and $\lambda=4$, become the
identities. By the explicit expressions for $\Gamma^{\pi}_{\rho\sigma}$ we write the
linear combinations of equations (\ref{23}), (\ref{24}):
\begin{equation}
\label{25} (u^3)^2(\frac{S_2}{r}-S_1\cot\theta)\sin\theta-mu^3-
\frac{MS_2}{r^4\sin\theta}(u^4)^2=-\frac{3MS_1}{r^3\cos\theta},
\end{equation}
$$
-m(u^3)^2\sin^2\theta+\frac{3u^3}{r^3}\left[S_2-S_1r\left(1-
\frac{3M}{r}\right)\tan\theta\right]\sin\theta+
$$
\begin{equation}
\label{26} +\frac{m}{r^2}\left(1-\frac{2M}{r}\right)(u^4)^2=0.
\end{equation}
Using relationship (\ref{22}), the set of second-order algebraic
equations (relative to $u^3$, $u^4$) (\ref{25}), (\ref{26}) can be
written as
\begin{equation}
\label{27} a_1(u^3)^2+a_2u^3+a_3(u^4)^2=0,
\end{equation}
\begin{equation}
\label{28} b_1(u^3)^2+b_2u^3+b_3(u^4)^2=d
\end{equation}
where
$$
a_1=1, \quad a_2=-\frac{6S_2}{mr^3\sin\theta}, \quad
a_3=-\frac{1-\frac{2M}{r}}{r^2\sin^2\theta},
$$
$$
b_1=\frac{S_2}{r\sin\theta}\left(1-\frac{3M}{r}\sin^2\theta\right)\left(1-
\frac{3M}{r}\right)^{-1}, \quad b_2=-m, \quad b_3=-\frac{MS_2}{r^4\sin\theta},
$$
\begin{equation}
\label{29} d=\frac{3MS_2}{r^4\sin\theta}\left(1-\frac{3M}{r}\right)^{-1}.
\end{equation}
In addition to equations (\ref{27}), (\ref{28}), the components of the particle's
4-velocity $u^3$, $u^4$ must satisfy the general relationship $u_\mu u^\mu=1$ which in
our case takes the form
\begin{equation}
\label{30} g_{33}(u^3)^2+g_{44}(u^4)^2=1.
\end{equation}

We point out that in (\ref{29}) $r\ne 3M$ because according to equations
(\ref{22})-(\ref{24}) at $r=3M$ we have the two possibilities: (i) $\cos\theta=0$, i.e.
the case of the equatorial circular motion which was considered in \cite{40}, and (ii)
$S_2=0, S_1\ne0$; however, it is easy to check that in the second case the expressions
for $u^3, u^4$ following from (\ref{25}), (\ref{26}) do not satisfy equation (\ref{30}).

Do equations (\ref{27}), (\ref{28}), (\ref{30}) have solutions at
$r\ne 3M$? To answer this question, let us substitute the
expression for $(u^4)^2$ through $(u^3)^2$ from (\ref{30}) in
(\ref{27}), (\ref{28}). Then we have
\begin{equation}
\label{31}
(u^3)^2(a_1-a_3\frac{g_{33}}{g_{44}})+a_2u^3+\frac{a_3}{g_{44}}=0,
\end{equation}
\begin{equation}
\label{32}
(u^3)^2(b_1-b_3\frac{g_{33}}{g_{44}})+b_2u^3+\frac{b_3}{g_{44}}-d=0.
\end{equation}
Taking into account (\ref{29}) we have $a_1-a_3g_{33}/g_{44}=0,$
and it follows from (\ref{31})
\begin{equation}
\label{33} u^3=-\frac{mr}{6S_2\sin\theta} \quad (S_2\ne 0).
\end{equation}
By (\ref{30}), (\ref{31}) we obtain
\begin{equation}
\label{34} u^4=\frac{mr^2}{6|S_2|}\left(1-\frac{2M}{r}\right)^{-1/2}\left(1+
\frac{36S_2^2}{m^2r^4}\right)^{1/2}.
\end{equation}
Substituting (\ref{33}) in (\ref{32}) we get the condition under
which three equations (\ref{27}), (\ref{28}), (\ref{30}) are
compatible
$$
\frac{m^2r^2}{36S_2^2\sin^2\theta}\left[\frac{S_2}{r\sin\theta}
\left(1-\frac{3M}{r}\sin^2\theta\right)\left(1-
\frac{3M}{r}\right)^{-1}-\frac{MS_2\sin\theta}{r^2}\left(1-
\frac{2M}{r}\right)^{-1}\right]
$$
\begin{equation}
\label{35} + \frac{m^2r}{6S_2\sin\theta}-\frac{MS_2}{r^4\sin\theta}\left(1-
\frac{2M}{r}\right)^{-1}-\frac{3MS_2}{r^4\sin\theta}\left(1- \frac{3M}{r}\right)^{-1}=0.
\end{equation}
Equation (\ref{35}) can be solved relative to $\sin^2\theta$:
$$
\sin^2\theta=\left(1-\frac{2M}{r}\right)\left[\frac Mr \left(4-\frac{9M}{r}\right)
\left(1+36\frac{S^2_2}{m^2r^4}\right)\right.
$$
\begin{equation}
\label{36} \left.-6\left(1-\frac{2M}{r}\right) \left(1-\frac{3M}{r}\right)\right]^{-1}.
\end{equation}

So, equation (1) in a Schwarzschild field at $r\ne 3M$ has the partial solutions with
(\ref{33}), (\ref{34}), (\ref{36}).

\subsection{The spatial region of existence of these solutions}

Now we shall write the relationship between $S_2$ and $S_0$. By
(\ref{9})-(\ref{11}), (\ref{20}), (\ref{22}) we find
\begin{equation}
\label{37} S^2_0=\frac{S^2_2}{r^2(u^4)^2}\left[1+\left(1-\frac{2M}{r}\right)
\left(1-\frac{3M}{r}\right)^{-2}\cot^2\theta\right]
\end{equation}
Taking into account expression (\ref{34}) we obtain from
(\ref{37})
$$
\frac{|S_0|}{mr}=\frac{6S^2_2}{m^2r^4}\left|-6\left(1-\frac{2M}{r}\right)
\left(1-\frac{3M}{r}\right)+36\frac Mr \left(4-\frac{9M}{r}\right)
\frac{S^2_2}{m^2r^4}\right|^{1/2}\times
$$
\begin{equation}
\label{38} \times\left[\left(1-\frac{2M}{r}\right)\left(1-\frac{3M}{r}\right)^2
\left(1+36\frac{S^2_2}{m^2r^4}\right)\right]^{-1/2}.
\end{equation}
We remind that the condition (\ref{8}) must be fulfilled.
Therefore, according to equations (\ref{8}) and (\ref{38}), it is
necessary
\begin{equation}
\label{39} \frac{6S^2_2}{m^2r^4}\ll 1.
\end{equation}
Then by (\ref{38}), (\ref{39}) we write
\begin{equation}
\label{40}
\frac{|S_0|}{mr}=\frac{6S^2_2}{m^2r^4}\sqrt{6}\left|1-\frac{3M}{r}\right|^{-1/2}.
\end{equation}

Naturally, the values of the right-hand side of equation
(\ref{36}) must be between 0 and 1. According to (\ref{36}) this
condition is fulfilled if
\begin{equation}
\label{41} \frac{15}{7}M\left(1+\frac 65\frac{S^2_2}{m^2r^4}\right)<r<
3M\left(1+\frac{6S^2_2}{m^2r^4}\right).
\end{equation}
By equation (\ref{39}) relationship (\ref{41}) can be written
approximately as
\begin{equation}
\label{42} \frac{15}{7} M<r<3M.
\end{equation}
This relationship determines the space region where equations
(\ref{1})-(\ref{3}) have the partial solutions for which
expressions (\ref{10}), (\ref{11}), (\ref{20}), (\ref{22}),
(\ref{33}), (\ref{34}), (\ref{36}), (\ref{42}) hold.

According to equation (\ref{36}) and (\ref{39}) the expression for $\sin^2\theta$ can be
written approximately as
$$
\sin^2\theta=\left(1-\frac{2M}{r}\right)\left[\frac Mr \left(4-\frac{9M}{r}\right)
\right.
$$
\begin{equation}
\label{43} \left.-6\left(1-\frac{2M}{r}\right) \left(1-\frac{3M}{r}\right)\right]^{-1}.
\end{equation}
By (\ref{43}) it is easy to find the interval of changes of
$\sin^2\theta$ when $r$ is changed in interval (\ref{42}). The
minimum value of $\sin^2\theta$ is
\begin{equation}
\label{44}\sin^2\theta|_{min}=0.465
\end{equation}
and this value is achieved at
\begin{equation}
\label{45} r=\frac{5\sqrt 5}{3\sqrt 5 -1}\approx 2.35M.
\end{equation}
At $r=2.35M$ and $r=2.5M$ we have from equation (\ref{43}) the
value $\sin^2\theta=0.5$, $\theta=45^0$. At $r=15M/7$ and $r=3M$
the value of $\sin^2\theta$ is maximum, namely $\sin^2\theta=1$.

\subsection{The interpretation of the above considered partial solutions
of equations (\ref{1})-(\ref{3})}

First, we stress that at $M=0$, i.e. in the Minkowski spacetime,
equation (\ref{23}) coincides with (\ref{24}) and we obtain
\begin{equation}
\label{46} u^3=\frac{mr\sin\theta}{S_2}.
\end{equation}
Naturally, in this case the restriction on the $\sin\theta$ does
not appear, because the geometric point $r=0$ and the plane
$\theta=\pi/2$ are not connected with any physical source.
Relationship (\ref{46}) describes the known motions of the
non-proper centers of mass of a spinning body.

It is important that expression (\ref{33}) which was obtained for
$M\ne 0$ does not pass into (\ref{46}) at $M\to 0$. This situation
is similar to the case for the equatorial circular orbits in a
Schwarzschild field which was investigated in \cite{40}
(relationship (\ref{46}) at $\theta=\pi/2$ corresponds to equation
(\ref{20}) from \cite{40}). Therefore, by the arguments discussed
in \cite{40} we conclude that expression (\ref{33}) describes the
motion of the proper center of mass of a spinning particle.

In this connection we point out the fact of importance. At
$r=15M/7$, when by (\ref{43}) $\theta=\pi/2$, according to
equation (\ref{22}) we have $S_1=0$, i.e. spin is orthogonal to
the equatorial plane of the particle motion. The expression for
the particle's orbital velocity for the equatorial circular orbits
with $2M<r<3M$ in a Schwarzschild field was written in \cite{40},
equation (24). It is easy to check that this expression coincides
with the corresponding expression following from equation
(\ref{33}) at $r=15M/7$, namely
$$
u^3=-\frac{5mM}{14S_2}.
$$

Using (\ref{8}), (\ref{33}), (\ref{34}), (\ref{40}) it is easy to
see that for the considered non-equatorial circular orbits the
relationships $(u^4)^2\gg 1$, $(ru^3)^2\gg 1$ take place. That is,
the velocity of a spinning particle on these orbits is
ultrarelativistic, as well as on the equatorial circular orbits
which were studied in \cite{40}.  

\subsection{The case of the shortened Mathisson-Papapetrou
equations}

In some papers, devoted to the investigations of the
Mathisson-Papapetrou equations, instead of strict equations
(\ref{1}) their shortened form was used \cite{6, 10}, namely
\begin{equation}\label{47}
m\frac D {ds} u^\lambda=-\frac {1} {2} u^\pi S^{\rho\sigma}
R^{\lambda}_{\pi\rho\sigma}.
\end{equation}
It means that condition (\ref{7}) is imposed on the
Mathisson-Papapetrou equations. As we pointed out in section 2,
the replacement of equations (\ref{1}) by equations (\ref{47})
eliminates the superfluous solutions of the strict
Mathisson-Papapetrou equations. It is important in this approach
that condition (\ref{7}) must be sutisfied.

In the context of our investigations it is interesting to verify
(i) do equations (\ref{2}), (\ref{47}) under condition (\ref{3})
have the non-equatorial circular solutions, and (ii) does
condition (\ref{7}) is satisfied on such possible solutions.
Therefore, in this subsection we shall consider equations
(\ref{47}), (\ref{2}), (\ref{3}) with relationships (\ref{10}),
(\ref{11}), (\ref{13}), (\ref{14}), (\ref{22}).

Two non-trivial equations of set (\ref{47}) with $\lambda=1$ and $\lambda=2$ are
\begin{equation}
\label{48} m(\Gamma^1_{33}u^3u^3+\Gamma^1_{44}u^4u^4)=-\frac{3M}{r^3}u^3S_2\sin\theta,
\end{equation}
\begin{equation}
\label{49} m\Gamma^2_{33}u^3u^3=-\frac{3M}{r^3}u^3S_1\sin\theta.
\end{equation}
From (\ref{49}) we have
\begin{equation}
\label{50} u^3=\frac{3MS_1}{mr^3\cos\theta}.
\end{equation}
By (\ref{48}), (\ref{50}), (\ref{22})
\begin{equation}
\label{51} (u^4)^2=\frac{9MS_1^2}{m^2r^5}\left(2-\frac{5M}{r}\right)
\left(1-\frac{2M}{r}\right)^{-1}\tan^2\theta.
\end{equation}
Substituting (\ref{50}), (\ref{51}) in condition (\ref{30}) we write
\begin{equation}
\label{52} \frac{18M}{r}\left(\frac{S_1}{mr}\right)^2\left(1-\frac{3M}{r}\right)
\tan^2\theta=1.
\end{equation}
According to (\ref{52}) it is necessary $r>3M$.

Using (\ref{9})-(\ref{11}), (\ref{20}), (\ref{22}) we find the relationship between $S_1$
and $S_0$
$$
S_0^2=S_1^2\left[\left(1-\frac{3M}{r}\right)^2\sin^2\theta+
\left(1-\frac{2M}{r}\right)\cos^2\theta\right]\times
$$
\begin{equation}
\label{53} \times\left(1-\frac{2M}{r}\right)^{-2}(u^4)^{-2}\cos^{-2}\theta.
\end{equation}
By (\ref{51}), (\ref{53})
$$
\frac{S_0^2}{m^2r^2}=\frac{r}{9M}\left[\left(1-\frac{3M}{r}\right)^2
\sin^2\theta+\left(1-\frac{2M}{r}\right)\cos^2\theta\right]\times
$$
\begin{equation}
\label{54} \times\left(1-\frac{2M}{r}\right)^{-1}\left(2-\frac{5M}{r}\right)^{-1}
\sin^{-2}\theta.
\end{equation}
Relationship (\ref{54}) shows that condition (\ref{8}) is
satisfied if and only if the value $r$ is close to $3M$ and
$\cos^2\theta$ is close to 0. Then equation (\ref{54}) can be
written as
\begin{equation}
\label{55} \frac{S_0^2}{m^2r^2}=3\delta_1^2+\delta_2^2
\end{equation}
where
$$
\delta_1\equiv 1-\frac{3M}{r}, \quad \delta_2\equiv \cos\theta,
$$
\begin{equation}
\label{56} 0<\delta_1\ll 1, \quad |\delta_2|\ll 1.
\end{equation}
By (\ref{56}) condition (\ref{52}) takes the form
\begin{equation}
\label{57} \frac{S_1^2}{m^2r^2}=\frac{\delta_2^2}{6\delta_1}.
\end{equation}
Taking into account (\ref{57}) we obtain from (\ref{50}), (\ref{51})
\begin{equation}
\label{58} (u^3)^2=\frac{1}{54M^2\delta_1}, \quad (u^4)^2=\frac{1}{2\delta_1}.
\end{equation}

So, shortened Mathisson-Papapetrou equations (\ref{47}), (\ref{2}) at condition (\ref{3})
have the non-equatorial circular solutions with the non-zero components of the particle
velocity determined by (\ref{58}). According to (\ref{56}), (\ref{58}) the relationships
$(u^4)^2\gg 1$, $(ru^3)^2\gg 1$ hold, i.e. the velocity of a spin particle on these orbit
is ultrarelativistic.

By the direct calculations it is not difficult to check that the
partial circular solutions of shortened Mathisson-Papapetrou
equations (\ref{2}), (\ref{3}), (\ref{47}) from this subsection
satisfy condition (\ref{7}) due to (\ref{56}).

\subsection{The value of the expression $S^{\mu\nu} P_\nu$ on the
non-equatorial circular orbits in a Schwarzschild field}

All partial solutions of the Mathisson-Papapetrou equations considered in this section
were obtained under supplementary condition (\ref{3}). At the same time the explicit
expressions above allow to estimate the value of the left-hand side of condition
(\ref{4}) on these solutions.

Using equations (\ref{22}), (\ref{22}), (\ref{33}), (\ref{34}), it
is easy to obtain the relationships
\begin{equation}
\label{59} S^{3\nu}P_\nu=0, \quad S^{4\nu}P_\nu=0.
\end{equation}
Other two expressions from the left-hand side of condition (\ref{4}), namely
$S^{1\nu}P_\nu$ and $S^{2\nu}P_\nu$, are not equal to 0. However, these expressions are
proportional to $\varepsilon$ (where $\varepsilon$ is determined by (\ref{8})). So, for
the small $\varepsilon$ the value $|S^{\mu\nu} P_\nu|$ is much less than 1. This fact may
be usefull for further investigations of the solutions of the Mathisson-Papapetrou
equations in a Schwarzschild field under condition (\ref{4}).

\subsection{Conclusions to section 3}

Thus, the strict Mathisson-Papapetrou equations under the Mathisson-Pirani supplementary
condition have the partial solutions describing the non-equatorial circular orbits of a
spinning test particle with constant latitude in a Schwarzschild field. The region of
existence of these orbits by the radial coordinate is determined by expression
(\ref{41}), or approximately by (\ref{42}). The dependence of the angle $\theta$ on $r$
is determined by expression (\ref{36}), or approximately by (\ref{43}). According to
(\ref{43}), for a spinning test particle hanging under the equatorial plane in a
Schwarzschild field the maximum value of the corresponding angle $\pi/2-\theta$ is equal
to $\approx 47^0$ and this value is achieved at $r\approx 2.35M$.

The shortened Mathisson-Papapetrou equations have in a
Schwarzschild field the solutions describing the non-equatorial
circular orbits as well. However, the region of existence of such
orbits is much less than in the case of the strict equations.
Namely, by (\ref{56}) $r$ is in the small neighborhood of $r=3M,$
and $\pi/2-\theta$ is close to 0.

All above considered non-equatorial orbits of a test spinning
particle in a Schwarzschild field are highly relativistic.

\section{The second-order Mathisson-Papapetrou equations for
non-circular equatorial motions in the Schwarzschild coordinates $r, \varphi$}

The equatorial ultrarelativistic circular orbits of a spinning
particle in a Schwarzschild field have been considered in
\cite{40}. Because any small perturbation deviates the orbit of a
spinning particle from the fixed circular orbit, it is interesting
to investigate the possible non-circular equatorial motions of
such a particle in a Schwarzschild field. Here we focus our
attention on same aspects of this question.

According to equation (10) from \cite{40}, for any equatorial
motions when spin is orthogonal to the motion plane the
relationship
\begin{equation}\label{60}
S_2=ru_4S_0
\end{equation}
holds ($S_1=0$, $S_3=0$). Taking into account (\ref{60}), we write
three nontrivial equations (\ref{1}) as
$$
S_0 r \left(1-\frac{2M}{r}\right)(u^3\ddot u^4-\ddot u^3 u^4)+
\dot u^1\left[m-3S_0u^3 u^4 \left(1-\frac{3M}{r}\right)\right]
$$
$$
-3S_0\left(1-\frac{2M}{r}\right) u^1 u^4\dot u^3+\frac{3S_0 M}{r}
u^1 u^3 \dot u^4 -\frac{3M}{r}\left(1-\frac{M}{r}\right)
$$
$$
\times \frac{S_0}{r}\left(1-\frac{2M}{r}\right)^{-1}(u^1)^2 u^3
u^4+rS_0 \left(1-\frac{2M}{r}\right)\left(1-\frac{3M}{r}\right)
(u^3)^3 u^4- \frac{S_0M}{r^2}
$$
\begin{equation}\label{61}
\times \left(1-\frac{2M}{r}\right)\left(1-\frac{3M}{r}\right)u^3
(u^4)^3+m\Gamma^1_{\alpha\beta} u^\alpha
u^\beta+\frac{3M}{r^2}\left(1-\frac{2M}{r}\right)S_0 u^3 u^4=0,
\end{equation}

$$
\frac{S_0}{r}(u^4\ddot u^1-\ddot u^4
u^1)-\frac{6S_0M}{r^3}\left(1-\frac{2M}{r}\right)^{-1} u^1 u^4\dot
u^1+\dot u^3\left[m-3S_0\left(1-\frac{2M}{r}\right) u^3 u^4\right]
$$
$$
-\frac{3S_0M}{r^3}\dot u^4\left[\left(1-\frac{2M}{r}\right)^{-1}
(u^1)^2-\left(1-\frac{2M}{r}\right)
(u^4)^2\right]+\frac{6MS_0}{r^4}\left(1-\frac{M}{r}\right)\left(1-\frac{2M}{r}\right)^{-2}
 (u^1)^3 u^4
$$
\begin{equation}\label{62}
-\frac{3S_0}{r}\left(1-\frac{2M}{r}\right) u^1 (u^3)^2
u^4-\frac{2S_0M}{r^4}\left(1-\frac{3M}{r}\right)
u^1(u^4)^3+2m\Gamma^3_{13} u^1 u^3=0,
\end{equation}

$$
S_0 r \left(1-\frac{2M}{r}\right)^{-1}(u^3\ddot u^1-\ddot u^3 u^1)
-3S_0\left(1-\frac{M}{r}\right)\left(1-\frac{2M}{r}\right)^{-2}u^1
u^3 \dot u^1-3S_0\dot u^3\left[r^2(u^3)^2\right.
$$
$$
\left.-\left(1-\frac{2M}{r}\right)^{-1}(u^1)^2\right] +\dot
u^4(m+\frac{3S_0 M}{r}u^3 u^4)+
\frac{3MS_0}{r^2}\left(1-\frac{M}{r}\right)\left(1-\frac{2M}{r}\right)^{-3}
(u^1)^3 u^3
$$
$$
-rS_0\left(1-\frac{2M}{r}\right)^{-1}\left(2-\frac{3M}{r}\right)u^1
(u^3)^3
-\frac{3S_0M}{r^2}\left(1-\frac{2M}{r}\right)^{-1}\left(1-\frac{3M}{r}\right)
$$
\begin{equation}\label{63}
\times u^1 u^3 (u^4)^2+\frac{3S_0M}{r^2}\left(1-\frac{2M}{r}\right)^{-1} u^1
u^3+2m\Gamma^4_{14} u^1 u^4=0.
\end{equation}
Using the condition $u_\mu u^\mu=1$ it is easy to check that among three equations
(\ref{61})-(\ref{63}) there are only two independent equations.

Further transformations of equations (\ref{61})-(\ref{63}) are connected with the
consideration of the known first integrals of the Mathisson-Papapetrou equations in a
Schwarzschild field, the energy $E$ and the angular momentum $L$ of a test particle
\cite{6}
$$
E=mu_4+g_{44} u_\mu \frac {DS^{4\mu}}{ds}+\frac12 g_{44,1}S^{14},
$$
\begin{equation}\label{64}
L=-mu_3-g_{33} u_\mu \frac {DS^{3\mu}}{ds}-\frac12 g_{33,1}S^{13}.
\end{equation}
By (\ref{60}) expressions (\ref{64}) can be written as
$$
E=m\left(1-\frac{2M}{r}\right)u^4-\frac{M}{r}S_0u^3+rS_0\left[\dot
u^1 u^3-u^1\dot u^3\right.
$$
\begin{equation}\label{65}
\left.+u^3\Gamma^1_{\alpha\beta}u^\alpha
u^\beta-2\Gamma^3_{13}(u^1)^2 u^3\right],
\end{equation}
$$
L=mr^2 u^3- \left(1-\frac{2M}{r}\right)S_0 u^4 - rS_0\left[\dot
u^4 u^1-u^4\dot u^1\right.
$$
\begin{equation}\label{66}
\left.+2\Gamma^4_{14}(u^1)^2 u^4-u^4\Gamma^1_{\alpha\beta}u^\alpha
u^\beta \right].
\end{equation}
From expressions (\ref{65}), (\ref{66}) we have
$$
\dot u^3=\dot
u^1\frac{u^3}{u^1}+\frac{u^3}{u^1}\Gamma^1_{\alpha\beta}u^\alpha
u^\beta -2\Gamma^3_{13}u^1 u^3-\frac{E}{rS_0u^1}
$$
\begin{equation}\label{67}
-\frac{M}{r^2}\frac{u^3}{u^1}+\frac{m}{rS_0}\frac{u^4}{u^1}
\left(1-\frac{2M}{r}\right),
\end{equation}
$$
\dot u^4=\dot
u^1\frac{u^4}{u^1}+\frac{u^4}{u^1}\Gamma^1_{\alpha\beta}u^\alpha
u^\beta - \Gamma^4_{\alpha\beta}u^\alpha u^\beta -
-\frac{L}{rS_0u^1}
$$
\begin{equation}\label{68}
+\frac{mr}{S_0}\frac{u^3}{u^1}-\frac{1}{r}\frac{u^4}{u^1}
\left(1-\frac{2M}{r}\right).
\end{equation}
Here we put $u^1\ne 0$ (the case of circular orbits was under
investigation in \cite{40}).

After the differentiation of equations (\ref{67}), (\ref{68}) with
respect to $s$ we obtain
$$
u^3\ddot u^1-\ddot u^3
u^1=-\frac{1}{u^1}\left[u^3\Gamma^1_{\alpha\beta}u^\alpha u^\beta
-u^1 \Gamma^3_{\alpha\beta}u^\alpha u^\beta-\frac{E}{rS_0}\right.
$$
$$
\left.-\frac{M}{r^2}
u^3+\frac{m}{rS_0}\left(1-\frac{2M}{r}\right)u^4\right]-\dot u^1
u^1
$$
$$
\times
\left[2u^3\left(\Gamma^1_{11}-\Gamma^3_{13}\right)+\frac{E}{rS_0(u^1)^2}\right]-
2\left[\dot u^1 u^3+u^3\Gamma^1_{\alpha\beta}u^\alpha
u^\beta\right.
$$
$$
\left. -u^1\Gamma^3_{\alpha\beta}u^\alpha
u^\beta-\frac{E}{rS_0}-\frac{Mu^3}{r^2}+\frac{m}{rS_0}\left(1-\frac{2M}{r}\right)u^4\right]
$$
$$
\times
\left(\frac{u^3}{u^1}\Gamma^1_{33}u^3-\Gamma^3_{13}u^1\right)-2\left[\dot
u^1 u^4- u^1\Gamma^4_{\alpha\beta}u^\alpha u^\beta +
u^4\Gamma^1_{\alpha\beta}u^\alpha u^\beta\right.
$$
$$
\left.
-\frac{L}{rS_0}+\frac{mru^3}{S_0}-\frac{1}{r}\left(1-\frac{2M}{r}\right)u^4\right]
\frac{u^3}{u^1}\Gamma^1_{44}u^4-u^1\Gamma^1_{\alpha\beta,1}u^3
u^\alpha u^\beta
$$
$$
+u^1\Gamma^3_{\alpha\beta,1}u^1 u^\alpha u^\beta-\frac{E}{r^2
S_0}u^1+\frac{m}{r^2 S_0}u^1\left(1-\frac{4M}{r}\right)u^1
u^4-\frac{2M}{r^3}u^1 u^4
$$
$$
-\frac{m}{u^1rS_0}\left(1-\frac{2M}{r}\right)\left[-u^1
\Gamma^4_{\alpha\beta}u^\alpha u^\beta+u^4
\Gamma^1_{\alpha\beta}u^\alpha u^\beta-\frac{L}{rS_0}\right.
$$
\begin{equation}\label{69}
 +\left.\frac{mru^3}{S_0}-\frac{1}{r}\left(1-\frac{2M}{r}\right)u^4\right].
\end{equation}
Taking into account equations (\ref{67})-(\ref{69}), we write equation (\ref{63}) in the
form
$$
\dot
u^1-\frac{1}{r}(u^1)^2-2r\left(1-\frac{3M}{r}\right)(u^3)^2-\frac{1}{r}
\left(1-\frac{3M}{r}\right)
$$
\begin{equation}\label{70}
+\frac{rE}{S_0}u^3-\frac{L}{rS_0}\left(1-\frac{2M}{r}\right)u^4=0.
\end{equation}
It is easy to check that equation (\ref{62}) by (\ref{67}),
(\ref{68}) coincides with equation (\ref{70}). From equation
(\ref{61}) according to (\ref{67}), (\ref{68}) we obtain
$$
\dot u^3+\frac{u^1
u^3}{r}-r\left(1-\frac{3M}{r}\right)\frac{(u^3)^2}{u^1}+\frac{E}{rS_0
u^1}\left[1+r^2(u^3)^2\right]
$$
\begin{equation}\label{71}
-\frac{1}{r}\left(1-\frac{3M}{r}\right)\frac{u^3}{u^1}-\frac{1}{ru^1S_0}(M+Lu^3)
\left(1-\frac{2M}{r}\right)u^4=0.
\end{equation}
Using the condition $u_\mu u^\mu=1$ we write equations (\ref{70}), (\ref{71}) in the
coordinates $r$ and $\varphi$:
$$
\ddot r=\frac{\dot r^2}{r}+2r\left(1-\frac{3M}{r}\right)\dot
\varphi^2-\frac{rE}{S_0}\dot \varphi
$$
\begin{equation}\label{72}
 +\frac{1}{r}\left(1-\frac{3M}{r}\right)+\frac{L}{rS_0}\left[\dot
 r^2+\left(1-\frac{2M}{r}\right)(1+r^2\dot \varphi^2)\right]^{1/2},
\end{equation}
$$
\ddot \varphi=-\frac{\dot r\dot \varphi}{r}+r\left(1-\frac{3M}{r}\right)\frac{\dot
\varphi^3}{\dot r}-\frac{E}{rS_0\dot r}(1+r^2\dot \varphi^2)
$$
\begin{equation}\label{73}
+\frac{m+L\dot \varphi}{rS_0\dot r}\left[\dot
 r^2+\left(1-\frac{2M}{r}\right)(1+r^2\dot \varphi^2)\right]^{1/2}.
\end{equation}
Thus, all possible equatorial world lines and trajectories (except the circular orbits)
of a spinning particle in a Schwarzschild field are described by equations (\ref{72}),
(\ref{73}).

We note that according to (\ref{64}) at $S_0\to 0$ we have
approximately $E=mu_4$, $L=-mu_3$, i.e. the expressions for the
geodesic motions. Substituting these values in (\ref{72}),
(\ref{73}) it is easy to see that the coefficients at the terms in
both these equations which are proportional to $1/S_0$ become
equal to 0. That is, at $S_0\to 0$ the right-hand sides in
(\ref{72}) and (\ref{73}) are finite. Besides, it is not difficult
to check that if the linear by spin corrections in expressions
(\ref{64}) are taken into account, then the right-hand sides of
equations (\ref{72}), (\ref{73}) coincide with the corresponding
expressions following from the equatorial geodesic equations.

\section{The concrete cases of highly relativistic essentially
 non-geodesic non-circular motions}

Here we shall confine ourselves to the cases of the non-circular
motions when the initial value of the particle's radial velocity
is much less then the initial value of its tangential velocity.
Such a choice is determined by the known fact that the
gravitational spin-orbit interaction decreases for the growing $r$
as $r^{-3}$. That is, just on the circular or close to circular
orbits a spinning particle feels the maximal effect of this
interaction.

We also point out that the ultrarelativistic circular orbits from \cite{40} are chosen as
the basic orbits in this section, i.e. the dynamics of the deviation of a spinning
particle from these orbits is under investigation.

\subsection{The equations in the non-dimensional quantities}

For further calculations, instead of equations
(\ref{72}),(\ref{73}) which are written in the variables $r(s)$
and $\varphi(s)$ we shall use their representation through the
non-dimensional quantities
\begin{equation}\label{74}
\tau\equiv\frac{s}{M},\quad Y\equiv\frac{dr}{ds},\quad Z\equiv M\frac{d\varphi}{ds},\quad
\rho\equiv\frac{r}{M}.
\end{equation}
Then from equations (\ref{72}), (\ref{73}) we have
$$
\frac{dY}{d\tau}=\frac{Y^2}{\rho}+\rho\left(1-\frac{3}{\rho}\right)\left(2Z^2+\frac{1}{\rho^2}\right)-\mu
Z\rho
$$
\begin{equation}\label{75}
+\frac{\nu}{\rho}\left[Y^2+\left(1-\frac{2}{\rho}\right)(1+Z^2\rho^2)\right]^{1/2},
\end{equation}
$$
\frac{dZ}{d\tau}=-\frac{YZ}{\rho}+\rho\frac{Z^2+1/\rho^2}{Y}\left(Z-\frac{3Z}{\rho}-\mu\right)
$$
\begin{equation}\label{76}
+ \frac{1}{\rho Y}\left(\frac{1}{\varepsilon} +\nu
Z\right)\left[Y^2+\left(1-\frac{2}{\rho}\right)(1+Z^2\rho^2)\right]^{1/2},
\end{equation}
\begin{equation}\label{77}
\frac{d\rho}{d\tau}=Y,
\end{equation}
\begin{equation}\label{78}
\frac{d\varphi}{d\tau}=Z
\end{equation}
where
\begin{equation}\label{79}
\mu=\frac{ME}{S_0},\qquad \nu=\frac{L}{S_0}.
\end{equation}
The small quantity $\varepsilon$ in (\ref{76}) is determined by
(\ref{8}). In the following we shall put $\varepsilon >0$, without
any loss in generality. Then according to (\ref{60}) $S_2>0$.

In the terms of quantities (\ref{74}), the condition that the
initial radial 4-velocity is much less by the absolute value than
the tangential 4-velocity can be written as
\begin{equation}\label{80}
|Y_0|\ll \rho_0 |Z_0|.
\end{equation}

Different values of the parameters $\mu$ and $\nu$ in equations (\ref{75}), (\ref{76}) at
the fixed initial values $\rho_0, \varphi_0, Y_0$ and $Z_0$ correspond to the solutions
describing motions of different centers of mass. It is important to pick out $\mu, \nu$
corresponding just to the proper center of mass. This question we shall consider below.

\subsection{The choice of the values $\mu$ and $\nu$ for $r_0=3M$}

First, we put that the initial value of the tangential velocity is
equal to the velocity of the spinning particle on the circular
orbit of radius $r=3M$ considered in \cite{40}. Then
\begin{equation}\label{81}
Z_0=-3^{-3/4}\varepsilon^{-1/2}\left(1-\frac{\sqrt{3}}{12}
\varepsilon+\frac{1}{96}\varepsilon^2\right).
\end{equation}
Using equations (\ref{64}) and notations (\ref{79}), by the direct
calculations we can obtain the values $\mu_c$ and $\nu_c$ for the
circular motion with $r=3M:$
\begin{equation}\label{82}
\mu_c=3^{-1/4}\varepsilon^{-3/2}\left(1+\frac{\sqrt{3}}{12}\varepsilon+\frac{1}{96}\varepsilon^2\right),
\end{equation}
\begin{equation}\label{83}
\nu_c=-3^{5/4}\varepsilon^{-3/2}\left(1-\frac{\sqrt{3}}{12}\varepsilon+\frac{1}{96}\varepsilon^2\right).
\end{equation}
According to the analysis from \cite{40}, relations (\ref{81})-(\ref{83}) correspond to
the motion of the proper center of mass of a spinning particle.

Naturally, in the cases of the non-circular motions, when $Y_0\ne0$,  the values $\mu$,
$\nu$ describing the motion of the proper center of mass differ from (\ref{82}),
(\ref{83}). However, because we are restricted by conditions (\ref{80}), (\ref{81}), i.e.
because we shall consider only the orbits close to the fixed circular orbit, we put
\begin{equation}\label{84}
\mu=k_1 \mu_c, \quad \nu=k_2 \nu_c
\end{equation}
where the coefficients $k_1$ and $k_2$ are close to 1. For the approximate estimates of
values $k_1, k_2$ we propose such a way. Let us consider the expressions for $dY/d\tau$
and $dZ/d\tau$ at the initial moment $\tau=0$. From equations (\ref{75}), (\ref{76}) by
(\ref{80})-(\ref{84}) we obtain
\begin{equation}\label{85}
\frac{dY}{d\tau}(0)=\frac{Y_0^2}{2}+\frac{1}{\varepsilon^2}
\left[k_1-k_2\left(1+\frac{Y_0^2\sqrt{3}}{2}\varepsilon\right)\right],
\end{equation}
$$
\frac{dZ}{d\tau}(0)=-\frac{3^{-3/4}}{Y_0\varepsilon^{5/2}}
\left(1-\frac{\sqrt{3}}{12}\varepsilon\right)\left[k_1-k_2\left(1+\frac{Y_0^2\sqrt{3}}{2}\varepsilon
\right)\right]
$$
\begin{equation}\label{86}
+\frac{3^{-5/4}}{Y_0\varepsilon^{3/2}}(1-k_1).
\end{equation}
According to notations (\ref{74}), expressions (\ref{85}), (\ref{86}) give the initial
values of the radial and tangential acceleration components. It is easy to see that the
right-hand sides of equations (\ref{85}), (\ref{86}) essentially depend on $k_1, k_2$ and
$\varepsilon$. It is important that the right-hand side of equation (\ref{85}) contains
the term with the large quantity $\varepsilon^{-2}$. In this connection we remember the
corresponding expressions following from the Mathisson-Papapetrou equations in the
Minkowski spacetime. Namely, it is known that in this case just the expressions for the
accelerations of the non-proper centers of mass are proportional to $(S_0/mr)^{-2}$, in
contrast to the acceleration of the proper center of mass which is equal to 0 (see, e.g.,
\cite{46, 50}; the value $S_0/mr$ in the Minkowski spacetime is analogous to
$\varepsilon$ from (\ref{8}) in a Schwarzschild spacetime). Here we can suppose that the
similar property posses the solutions of the Mathisson-Papapetrou equations in a
Schwarzschild spacetime. Namely, that the term in (\ref{85}) which is proportional to
$\varepsilon^{-2}$ determines the motions of the non-proper centers of mass, whereas for
the description of the proper center of mass it is necessary to put
\begin{equation}\label{87}
k_1-\left(1+\frac{Y_0^2\sqrt{3}}{2}\varepsilon\right)k_2 \approx 0,
\end{equation}
i.e. to decrease the effect of the term with $\varepsilon^{-2}$ on the initial value of
the radial acceleration.

We stress that the reflection above concerning the choice of $\mu,
\nu$ for the proper center of mass is not strict (its role is
heuristic) and must be verified in the concrete computer
calculations. We shall do it in the next subsection.

It is easy to see that under condition (\ref{87}) the effect of the large quantity
$\varepsilon^{-5/2}$ on the tangential acceleration in equation (\ref{86}) is decreased
as well. Besides at the condition
\begin{equation}\label{88}
k_1\approx 1
\end{equation}
the influence of the large quantity $\varepsilon^{-3/2}$ in
(\ref{86}) is decreased. Because on the circular orbit
$dY/d\tau=0, dZ/d\tau=0$, it is naturally to suppose that at
condition (\ref{80}) expressions (\ref{85}), (\ref{86}) cannot be
too large in the case of the proper center of mass.

Therefore, according to relationships (\ref{87}), (\ref{88}), in computer integrations of
equations (\ref{75})-(\ref{78}) we start with the expressions
\begin{equation}\label{89}
k_1 = 1, \quad
k_2=\left(1+\frac{Y_0^2\sqrt{3}}{2}\varepsilon\right)^{-1}.
\end{equation}

\subsection{The results of the computer integration of equations (\ref{75})-(\ref{78}) at $r_0=3M$}

Typical results of the integration of equations (\ref{75})-(\ref{78}) at (\ref{84}),
(\ref{89}) are summarized in figures 1, 2 in terms of the radial velocity $dr/ds=Y$, the
value $r/M=\rho$ and the angle $\varphi$ at $\varepsilon=10^{-4}$. By (\ref{81}) for this
value $\varepsilon$ we have $Z_0\approx -44$. In figures 1, 2 the interval of integration
of equations (\ref{75})-(\ref{78}) by $\tau$ is [0--0.2]. According to figure 1 in this
interval the angle $\varphi$ is changed from $0$ to $-8$ radians that corresponds to
$\approx 1.3$ revolution around the Schwarzschild mass. The plot of $\varphi(\tau)$ is
not presented in figure 2 because it practically coincides with the corresponding plot in
figure 1. The plots of $Z(\tau)$ is absent in figures 1, 2 because in the considered
interval of $\tau$ the value $Z$ is practically constant.

Figure 3 shows the solutions of equations (\ref{75})-(\ref{78}) in the case when the
first relation from (\ref{89}) is violated, namely at $k_1= 1+2\times 10^{-7}$, whereas
the second relation from (\ref{89}) remains the same. The oscillation motion of the
non-proper center of mass is visible in figure 3. More generally, the computer
integration of equations (\ref{75})-(\ref{78}) shows that the similar situation takes
place for other concrete values $k_1, k_2$ which violate condition (\ref{89}). In this
connection we conclude that relation (\ref{89}) are correct for the description of the
motion just of the proper center of mass.

In subsection 5.7 we shall discuss the results of subsection 5.3.

\begin{figure}[h]
\begin{center}
\includegraphics[height=5cm]{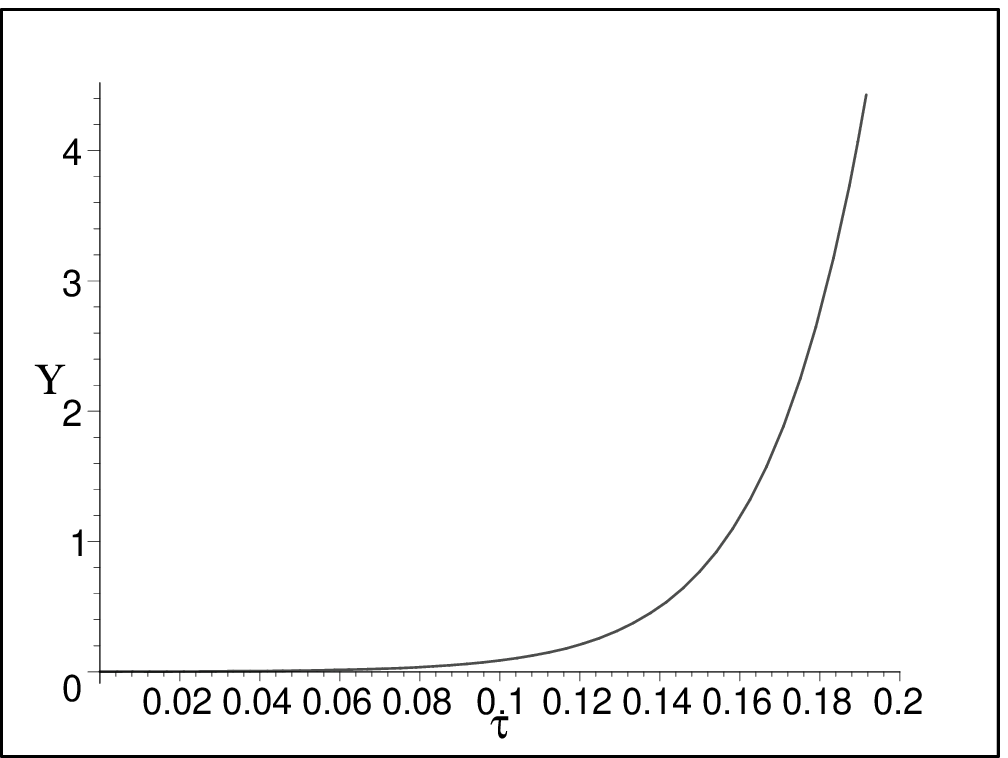}
\includegraphics[height=5cm]{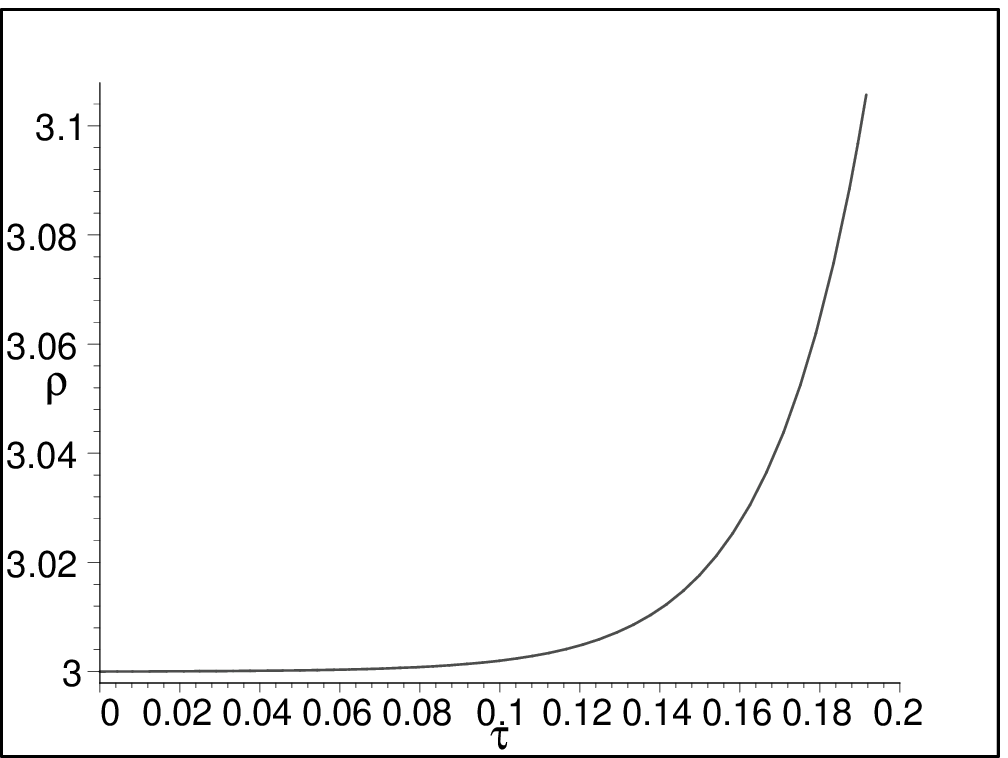}
\includegraphics[height=5cm]{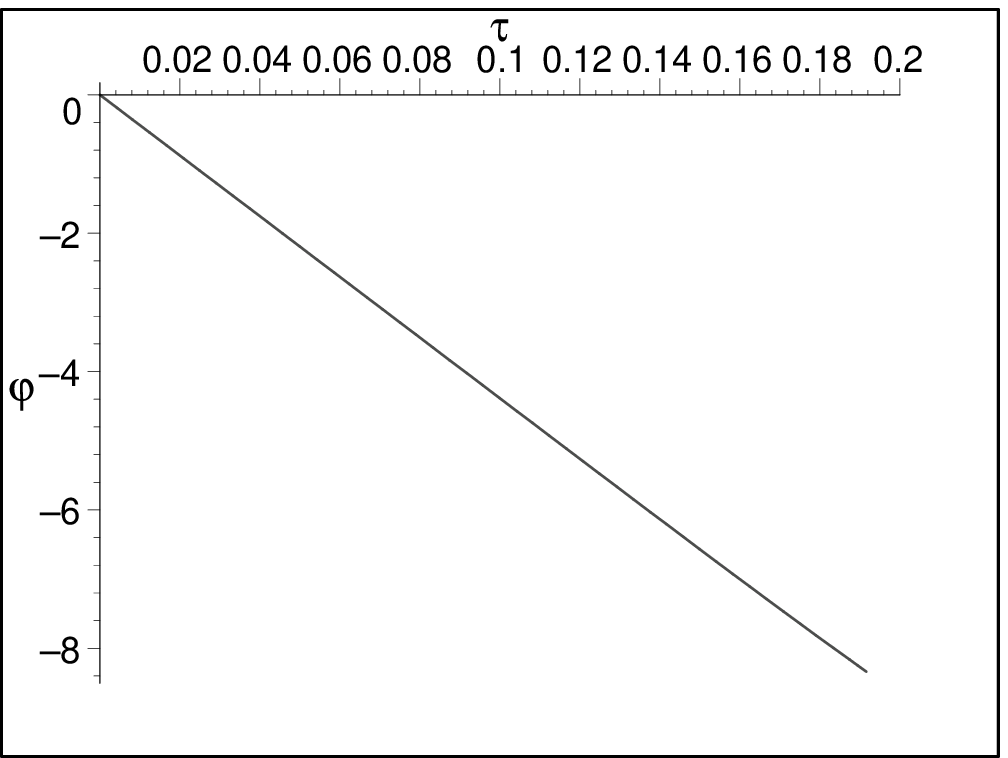}
\end{center}
\caption{ $Y\equiv dr/ds, \rho\equiv r/M$ and $\varphi$ vs. $\tau$
by the strict Mathisson-Papapetrou (MP) equations for
$\varepsilon=10^{-4}, \rho_0=3, Y_0=2.5\times10^{-3}.$
}\label{Figure 1}
\end{figure}

\begin{figure}[h]
\begin{center}
\includegraphics[height=5cm]{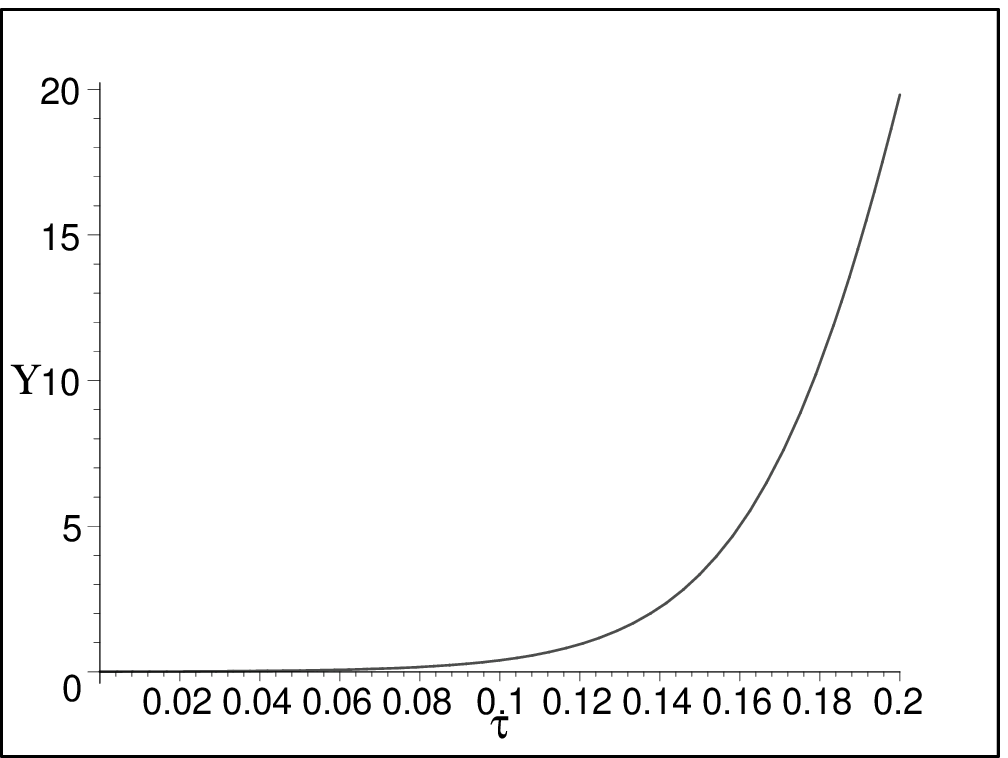}
\includegraphics[height=5cm]{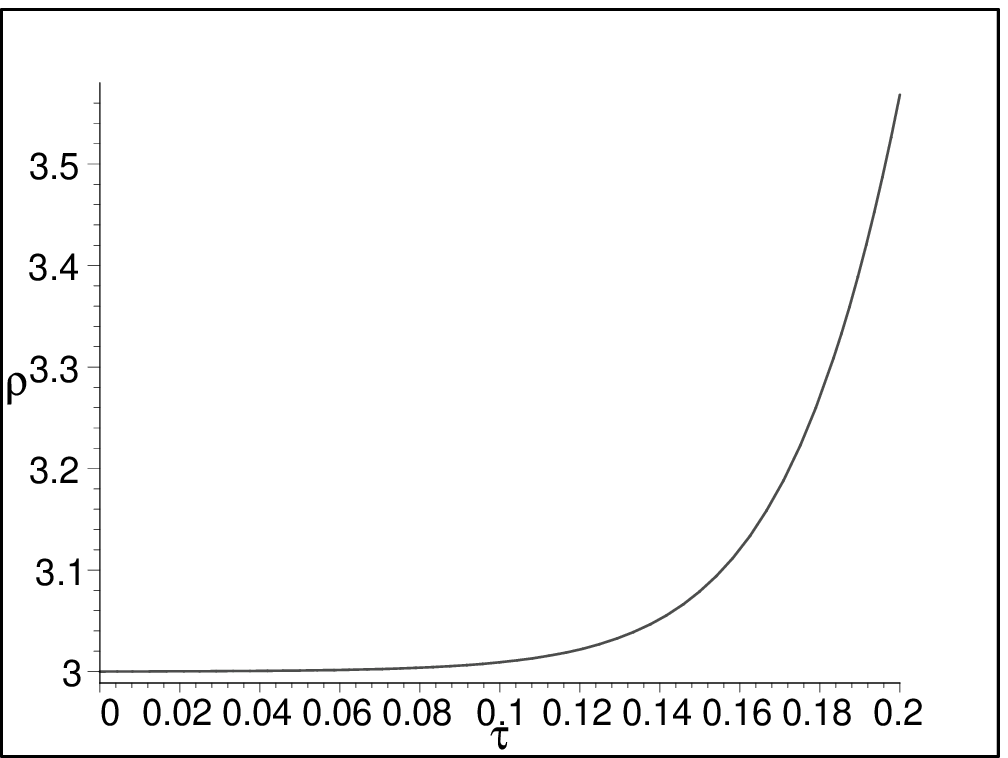}
\end{center}
\caption{$Y$ and $\rho$ vs. $\tau$ by the strict MP equations for $\varepsilon=10^{-4},
\rho_0=3, Y_0=10^{-2}.$ }\label{Figure 2}
\end{figure}

\newpage
\begin{figure}[h]
\begin{center}
\includegraphics[height=5cm]{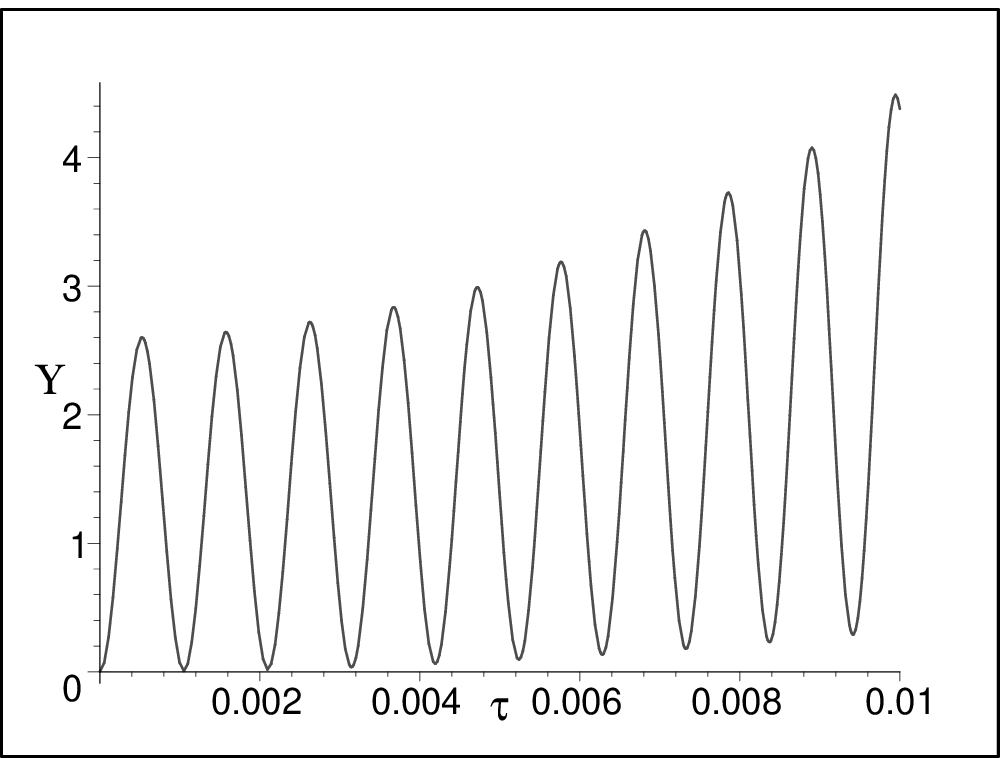}
\includegraphics[height=5cm]{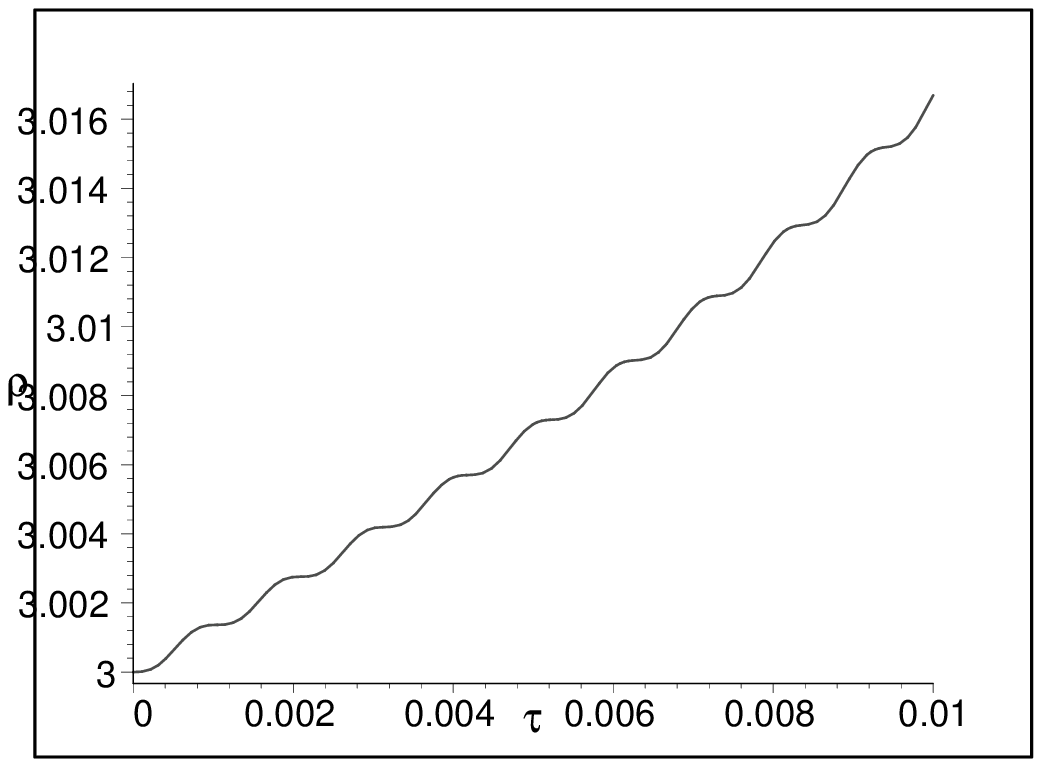}
\end{center}
\caption{The examples of the oscillation solutions of $Y$ and $\rho$ by the strict MP
equations for $\varepsilon=10^{-4}, \rho_0=3, Y_0=2.5\times10^{-3}.$ }\label{Figure 3}
\end{figure}


\subsection{The results of the computer integration of equations
(\ref{75})-(\ref{78}) at $r_0=2.5M$}

It is pointed out in \cite{40} that in a Schwarzschild field the ultrarelativistic
circular orbits of the proper center of mass of the spinning particle are admissible at
$2M<r<3M$ as well. In contrast to the circular orbit of radius $r=3M$, which follows both
from the strict and shortened Mathisson-Papapetrou equations, the orbits with $r<3M$
follow from the strict Mathisson-Papapetrou equations only. Similarly to the above
considered deviation of a spinning particle from the orbit $r=3M$, it is possible to
study the deviation of a particle from the orbit with $r<3M$. The initial equations are
the same, namely (\ref{75})-(\ref{78}).

As a typical case, let us consider ultrarelativistic motions which deviate from the
circular orbit of radius $r=2.5M$. By the results of \cite{40}, we found for this
circular orbit
\begin{equation}\label{90}
Z_0=-2^{1/2}5^{-1/4}\varepsilon^{-1/2}\left(1-\frac{13\sqrt{5}}{50}\varepsilon-0.023\varepsilon^2\right),
\end{equation}
\begin{equation}\label{91}
\mu_c=7\cdot 5^{-3/4}\sqrt{2}\varepsilon^{-1/2}\left(1-\frac{23
\sqrt{5}}{70}\varepsilon\right),
\end{equation}
\begin{equation}\label{92}
\nu_c=-\frac{5^{5/4}}{\sqrt{2}}\varepsilon^{-1/2}\left(1+\frac{397\sqrt{5}}{250}\varepsilon\right)
\end{equation}
(the corresponding expressions for the circular orbit with
$\rho=3$ were written in (\ref{81})-(\ref{83})).

By the procedure similar to the described above (see equations (\ref{84})-(\ref{89})), we
can obtain the expressions for $\mu, \nu$ corresponding to the proper center of mass when
$Y_0\ne 0$ at condition (\ref{80}):
\begin{equation}\label{93}
\mu=\mu_c\left[1-\frac{100}{821}Y_0^2(1+87.5Y_0^2)^{-1}\right],
\end{equation}
\begin{equation}\label{94}
\nu=\nu_c\left[1-\frac{140}{821}Y_0^2(1+87.5Y_0^2)^{-1}\right].
\end{equation}

Figure 4 shows typical results of the integration of equations
(\ref{75})-(\ref{78}) at conditions (\ref{93})-(\ref{94}) for
$\rho_0=2.5$. These results are discussed in subsection 5.7.
\begin{figure}[h]
\begin{center}
\includegraphics[height=5cm]{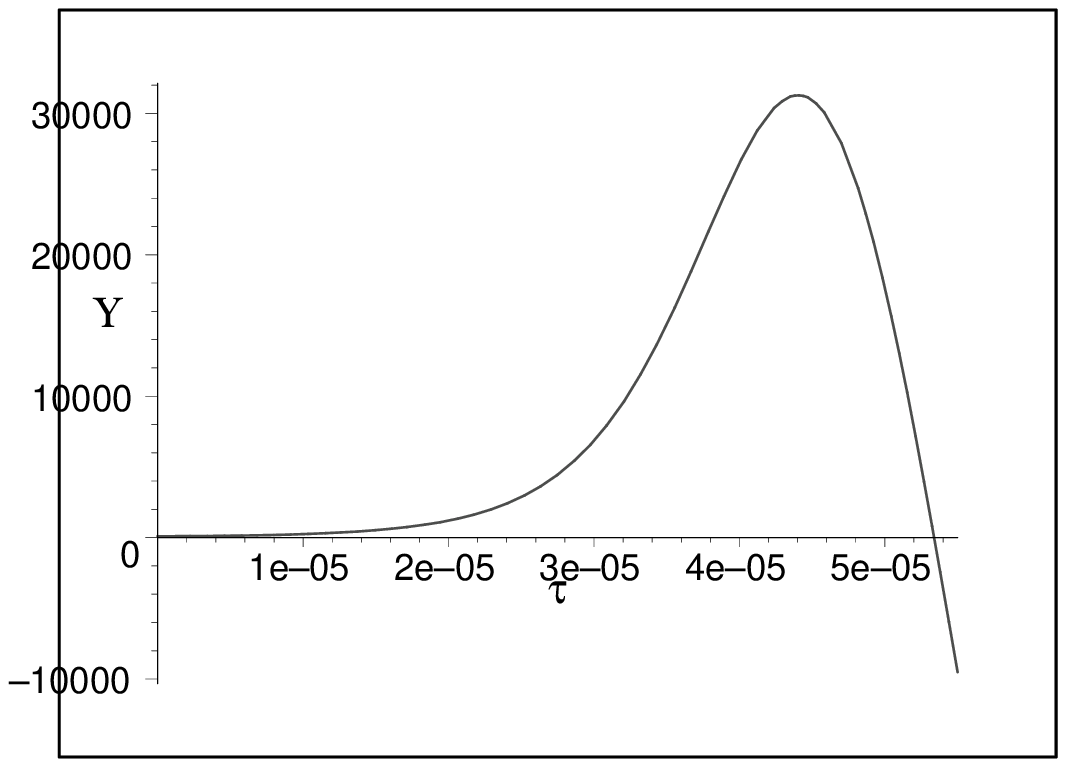}
\includegraphics[height=5cm]{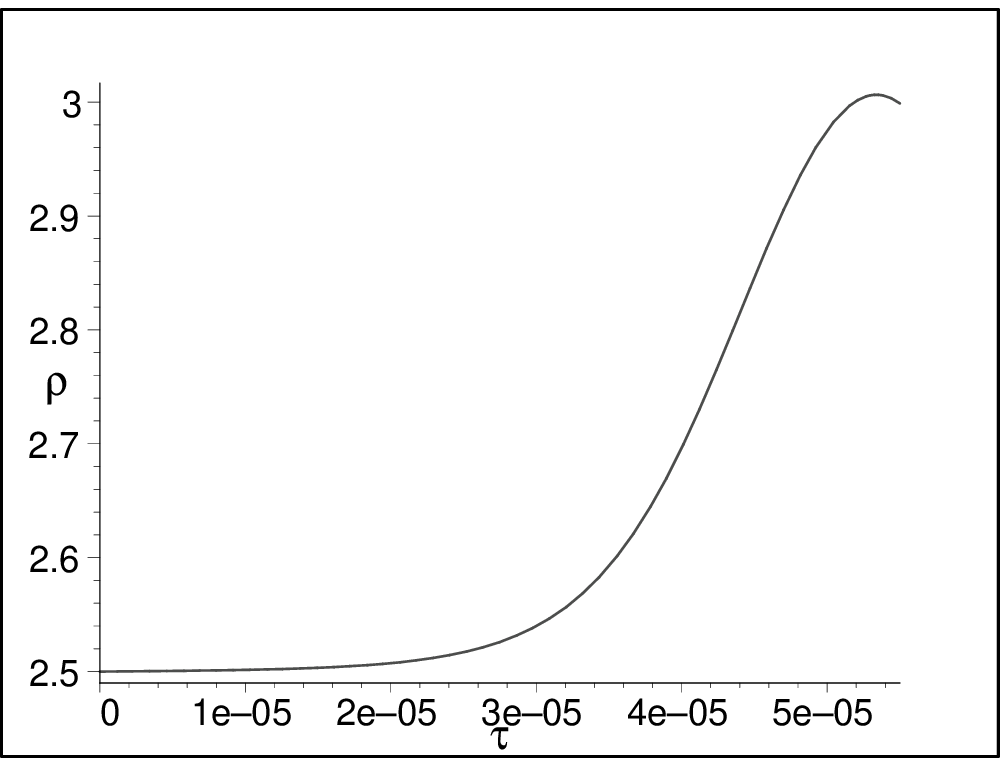}
\includegraphics[height=5cm]{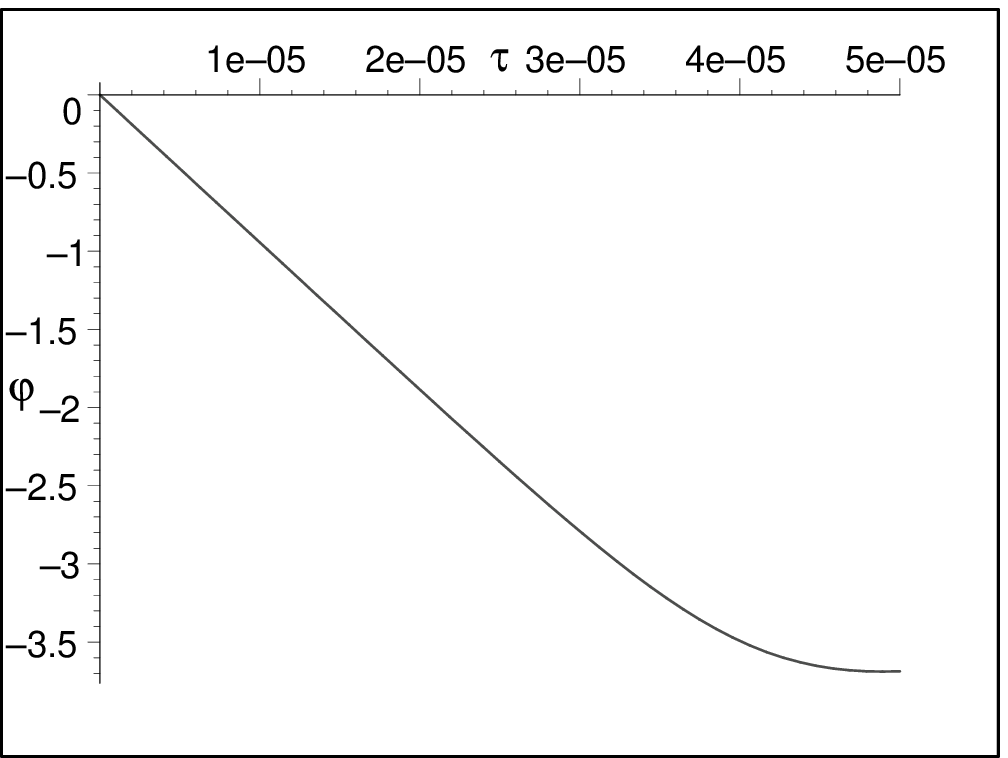}
\end{center}
\caption{$Y, \rho$ and $\varphi$ vs. $\tau$ by the strict MP equations for
$\varepsilon=10^{-10}, \rho_0=2.5, Y_0=100.$ }\label{Figure 4}
\end{figure}

\subsection{The case of the shortened Mathisson-Papapetrou equations}

For the equatorial motions of a spinning particle in a Schwarzschild field equations
(\ref{47}) can be written in the coordinates $r, \varphi$ as
\begin{equation}\label{95}
\ddot r=\dot\varphi^2 r\left(1-\frac{3M}{r}\right)-\frac{M}{r^2}- \frac{3MS_0}{r^2
m}\dot\varphi\left[\dot
 r^2+\left(1-\frac{2M}{r}\right)(1+r^2\dot
 \varphi^2)\right]^{1/2},
\end{equation}
\begin{equation}\label{96}
\ddot \varphi=-\frac{2}{r}\dot r\dot\varphi.
\end{equation}
At $S_0=0$ these equations coincide with the geodesic equations.

In quantities (\ref{74}) equations (\ref{95}), (\ref{96}) can be written as
$$
\frac{dY}{d\tau}=Z^2\rho\left(1-\frac{3}{\rho}\right)-\frac{1}{\rho^2}
$$
\begin{equation}\label{97}
-3\varepsilon\frac{Z}{\rho^2}\left[Y^2+\left(1-\frac{2}{\rho}\right)(1+Z^2\rho^2)\right]^{1/2},
\end{equation}
\begin{equation}\label{98}
\frac{dz}{d\tau}=-2\frac{YZ}{\rho},
\end{equation}
\begin{equation}\label{99}
\frac{d\rho}{d\tau}=Y,
\end{equation}
\begin{equation}\label{100}
\frac{d\varphi}{d\tau}=Z.
\end{equation}
An example of the integration of equations (\ref{97})-(\ref{100}) is presented in figure
5 .

\begin{figure}[h]
\begin{center}
\includegraphics[height=5cm]{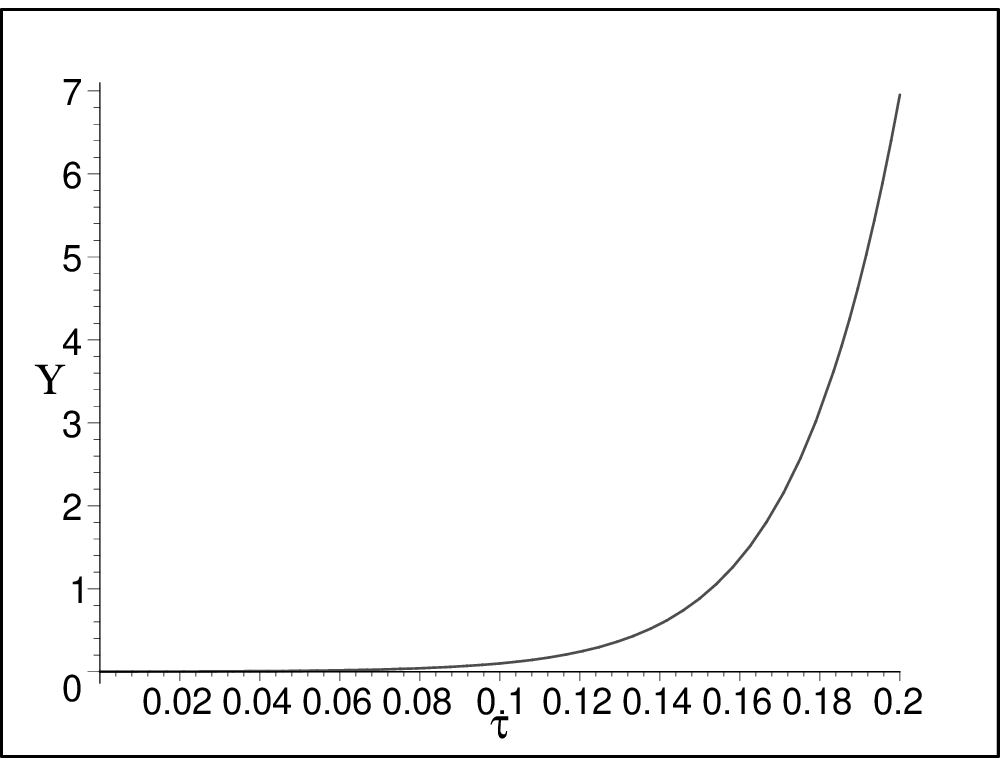}
\includegraphics[height=5cm]{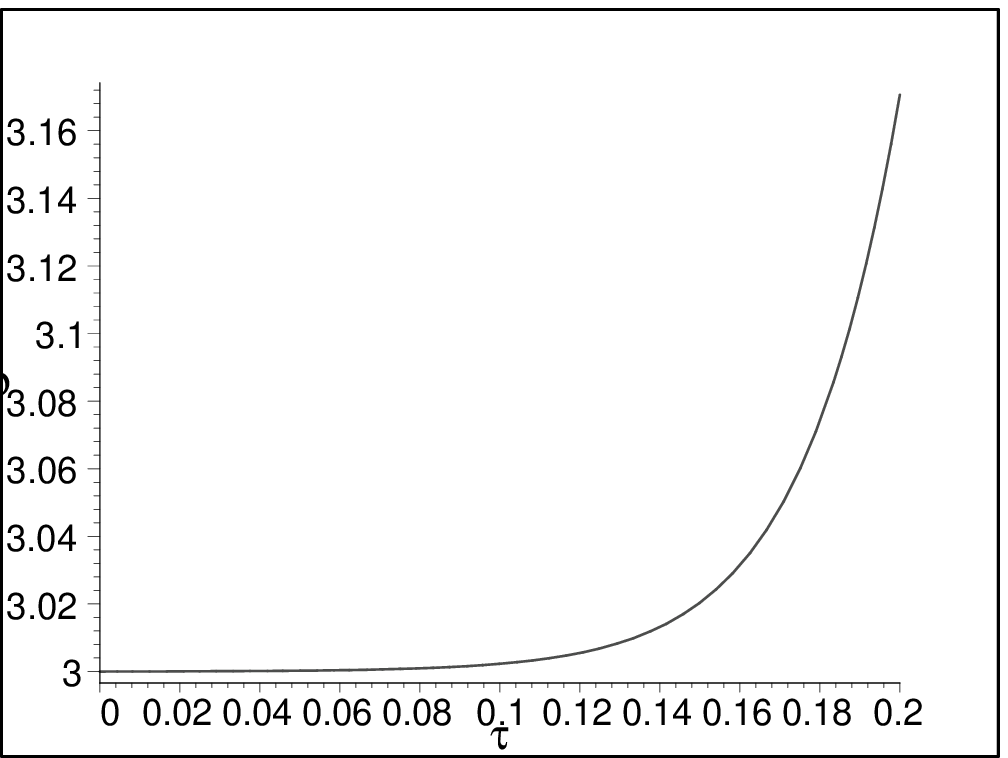}
\end{center}
\caption{$Y$ and $\rho$ and  vs. $\tau$ by the shortened MP equations for
$\varepsilon=10^{-4}, \rho_0=3, Y_0=2.5\times10^{-3}.$ }\label{Figure 5}
\end{figure}
\subsection{The case of the geodesic equations}

In notations (\ref{74}) the geodesic equations coincide with equations
(\ref{97})-(\ref{100}) when $\varepsilon$ in (\ref{97}) is equal to 0. It is interesting
to compare the corresponding solutions of the Mathisson-Papapetrou and geodesic
equations. Figures 6--8 we shall use for the discussion of the results following from
figures 1, 2, 4.

\newpage
\begin{figure}[t]
\begin{center}
\includegraphics[height=5cm]{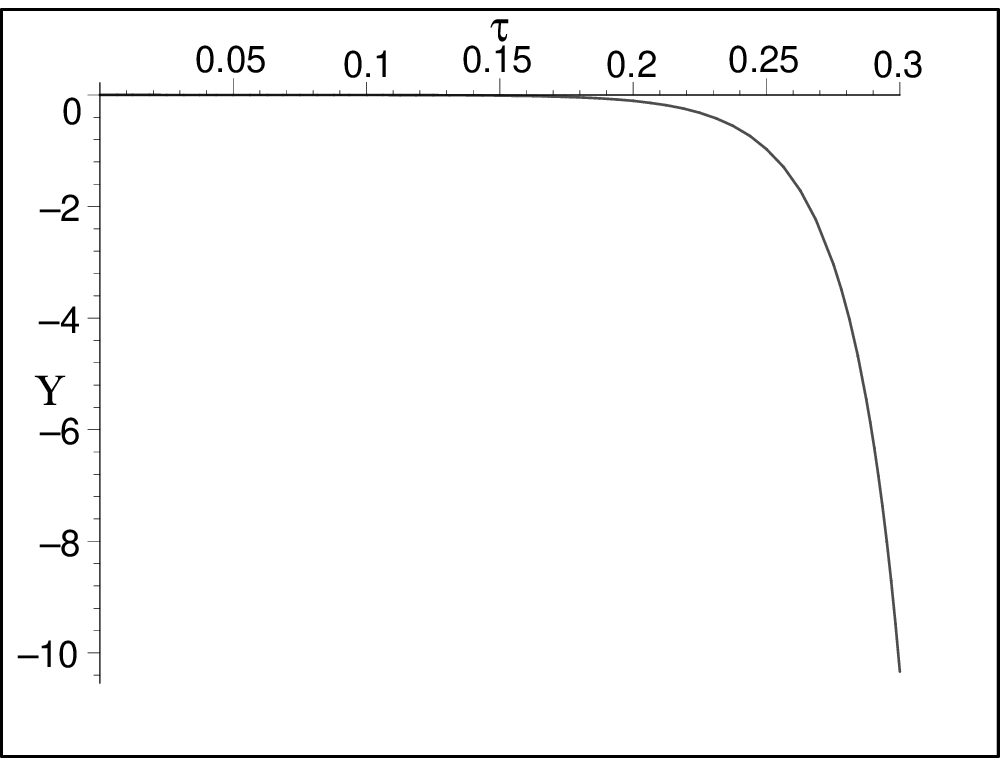}
\includegraphics[height=5cm]{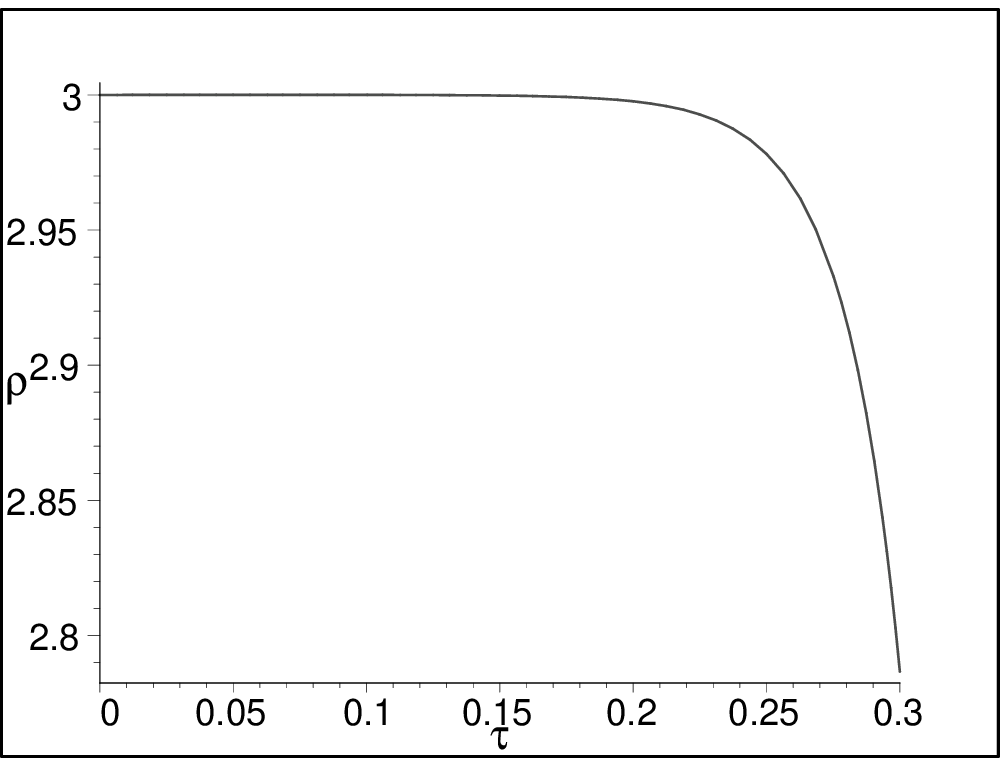}
\includegraphics[height=5cm]{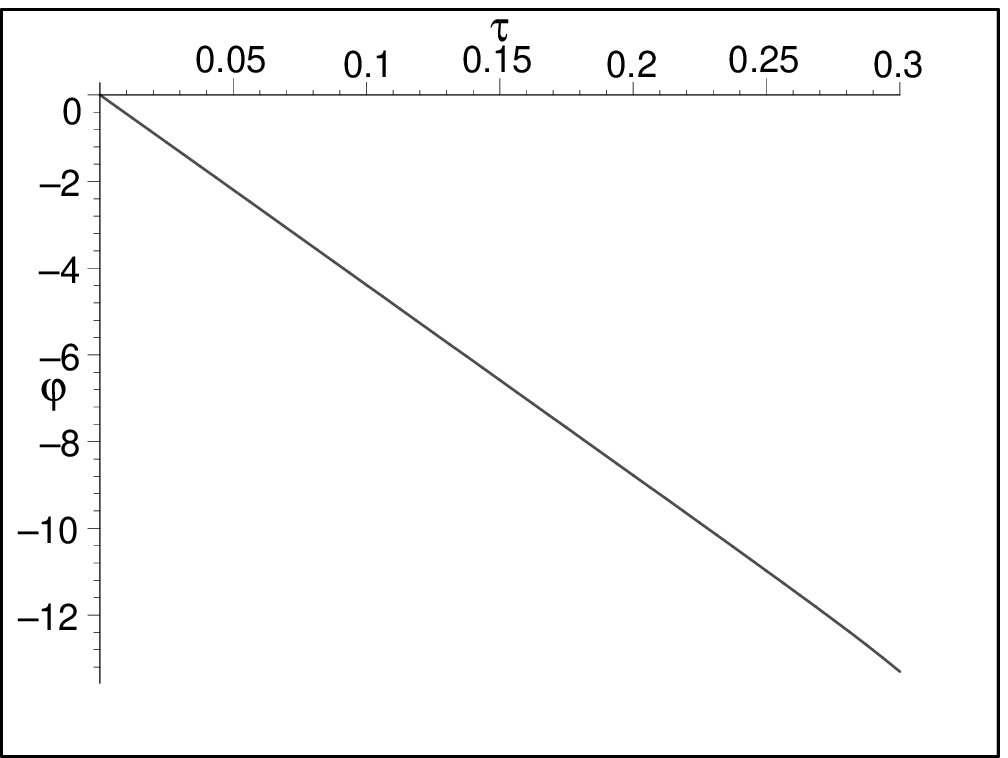}
\end{center}
\caption{$Y, \rho$ and $\varphi$ vs. $\tau$ by the geodesic equations for
$\varepsilon=10^{-4}, \rho_0=3, Y_0=2.5\times10^{-3}.$ }\label{Figure 6}
\end{figure}

\begin{figure}[h]
\begin{center}
\includegraphics[height=5cm]{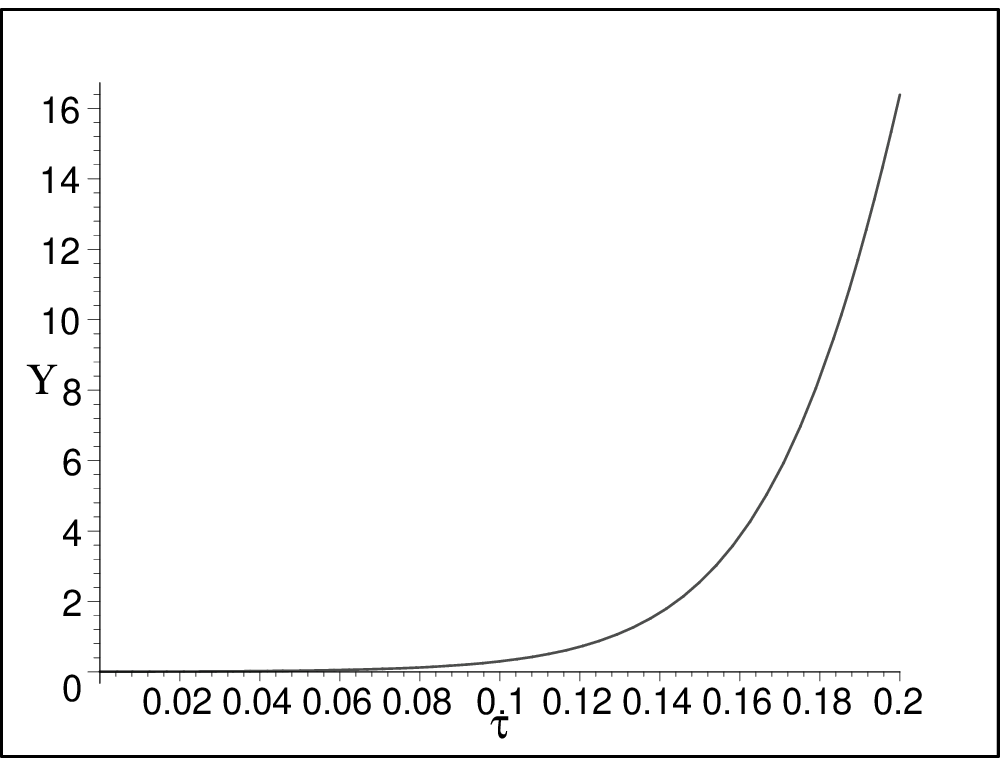}
\includegraphics[height=5cm]{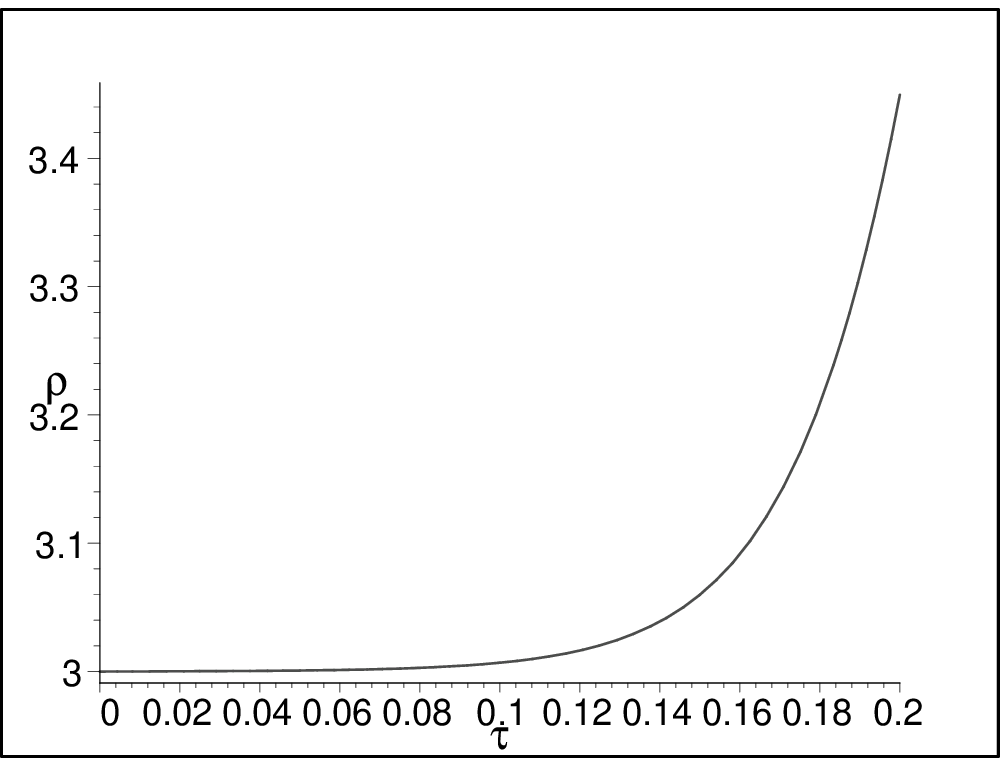}
\end{center}
\caption{$Y$ and $\rho$ vs. $\tau$ by the geodesic equations for $\varepsilon=10^{-4},
\rho_0=3, Y_0=10^{-2}.$}\label{Figure 7}
\end{figure}

\begin{figure}[h]
\begin{center}
\includegraphics[height=5cm]{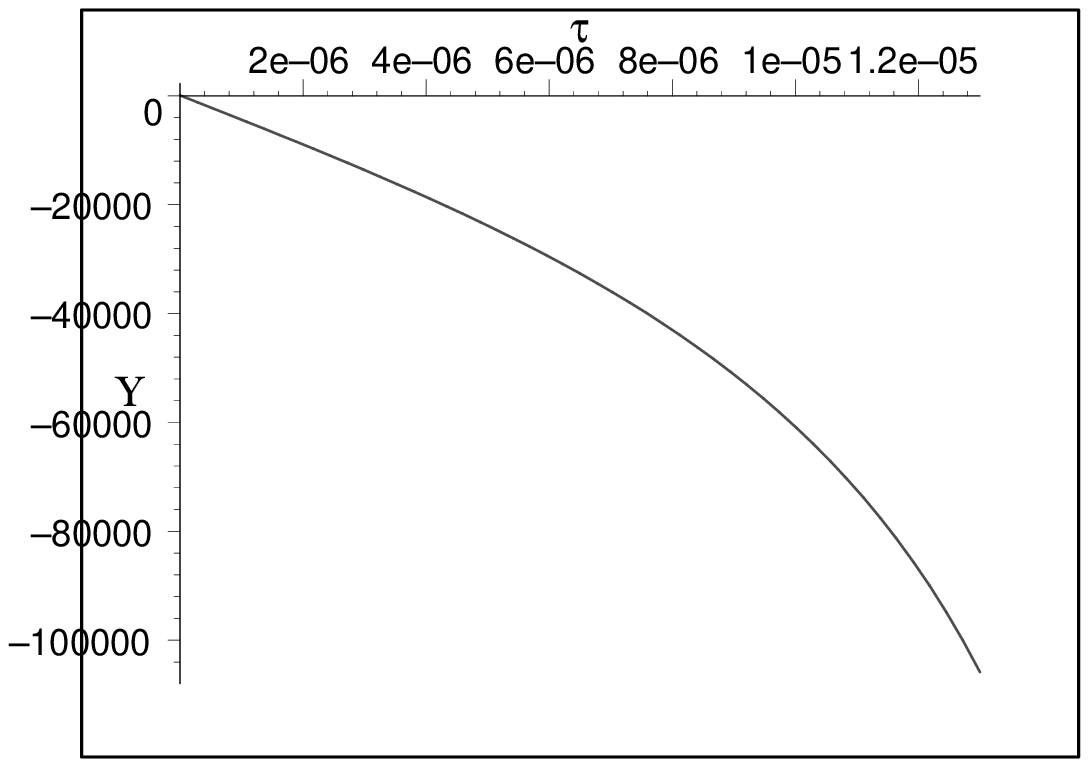}
\includegraphics[height=5cm]{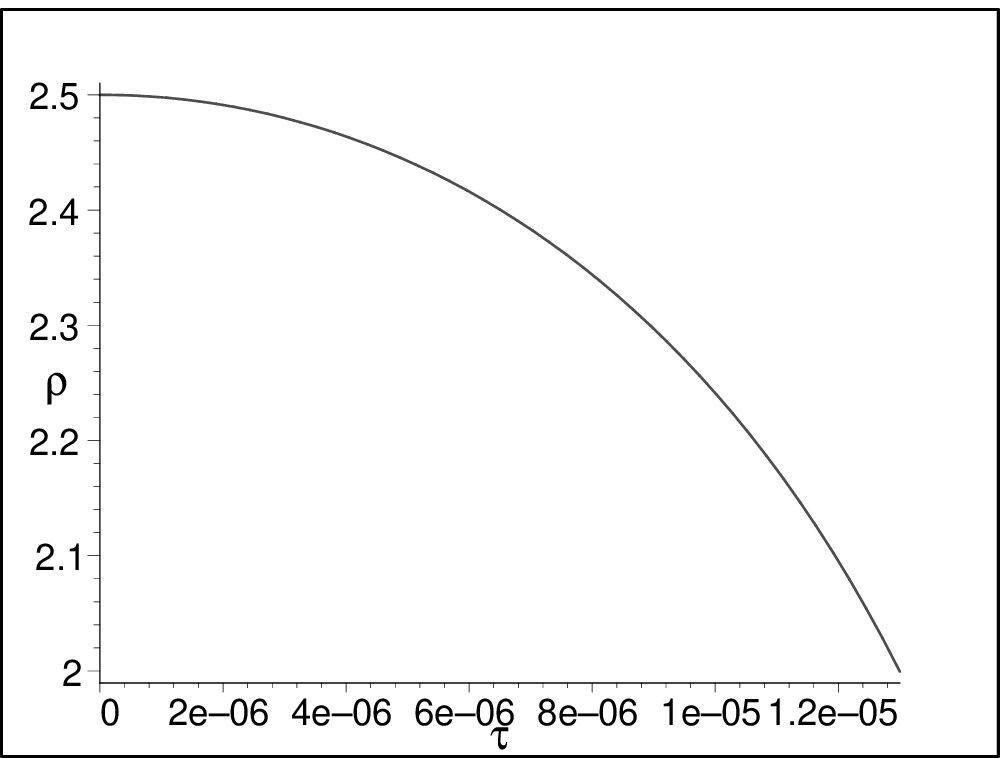}
\includegraphics[height=5cm]{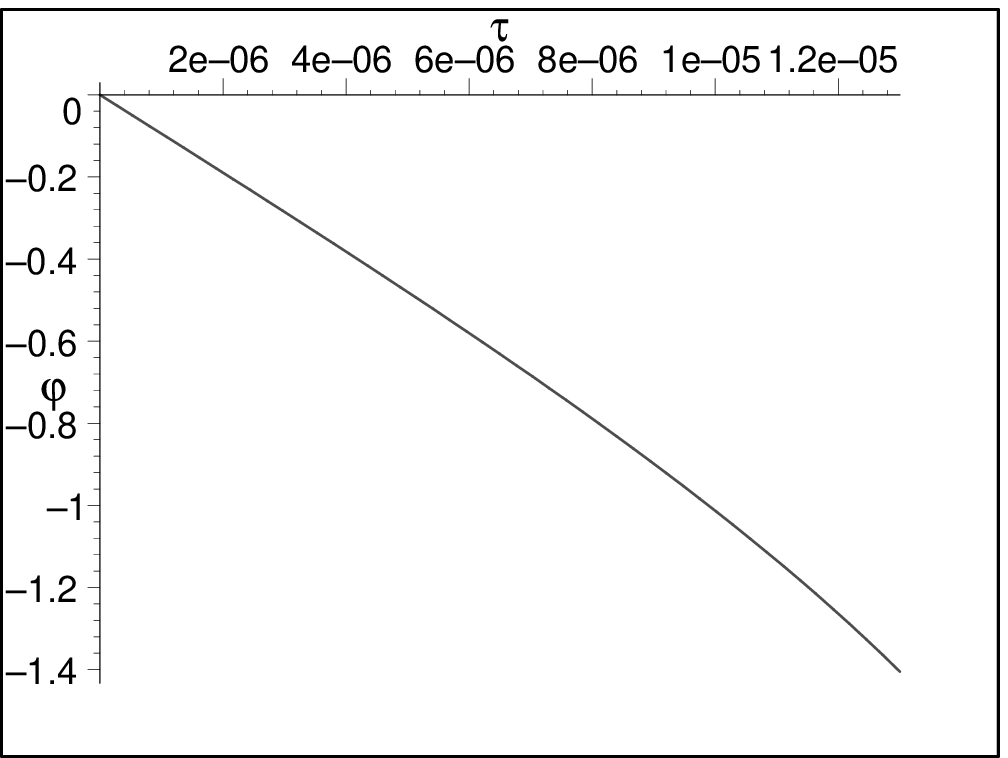}
\end{center}
\caption{$Y, \rho$ and $\varphi$ vs. $\tau$ by the geodesic equations for
$\varepsilon=10^{-10}, \rho_0=2.5, Y_0=100.$ }\label{Figure 8}
\end{figure}

\newpage
\subsection{Discussion}

First, let us compare figures 1, 5 and 6 which show the corresponding solutions of the
strict Mathisson-Papapetrou equations (\ref{75})-(\ref{78}), the shortened equations
(\ref{97})-(\ref{100}) and the geodesic equations respectively, with the same initial
values of $\rho_0, Y_0,Z_0$: $\rho_0=3, Y_0=2.5\times10^{-3}, Z_0=-44,$ at
$\varepsilon=10^{-4}$. As figures 1 and 5 show, on the $\tau$-scale of order $0.2$ the
corresponding plots of $Y$ and $\rho$ are closer. (The plot of $\varphi(\tau)$ is not
presented in figure 5 because it is not distinguished from $\varphi(\tau)$ in figure 1).
Whereas the plots $Y(\tau)$ and $\rho(\tau)$ in figure 6 essentially differ from the
corresponding plots in figure 1: according to figure 1 the spinning particle moves away
from a Schwarzschild mass, while by figure 6 the spinless particle falls onto a
Schwarzschild mass. We stress that the difference of the radial coordinate of both the
particles $\Delta r$ becomes comparable with $r$ after less than 2 revolutions around a
Schwarzschild mass.

The initial values in figures 2 and 7 are: $\rho_0=3, Y_0=10^{-2}, Z_0=-44,$ at
$\varepsilon=10^{-4}$. These figures show the situation when both the spinning particle
and the spinless particle move away from the Schwarzschild mass but with different
velocities.

Let us compare figures 4 and 8. Here $\varepsilon=10^{-10}, \rho_0=2.5, Y_0=100$ and, by
expression (\ref{90}), $Z_0=-9.5\times10^4$. According to figure 4 for the $\tau$-scale
of order $5\times10^{-5}$ the spinning particle moves away from a Schwarzschild mass,
reaches the value approximately $\rho=3$ and after that moves toward a Schwarzschild
mass. Whereas according to figure 8 the spinless particle monotonously falls onto a
Schwarzschild mass and for the $\tau=1.2\times10^{-5}$ reaches the horizon surface
$\rho=2$. We point out that the plots of $\varphi(\tau)$ in figures 4 and 8 show that
these processes correspond to less than one particle's revolution around a Schwarzschild
mass.

\section{Conclusions}

The analytical solutions of the Mathisson-Papapetrou equations
from section 3 and the numerical solutions from section 5 have a
common feature: they describe such a type of ultrarelativistic
motions of a spinning particle when the interaction of spin with
the gravitational field acts as a repulsive force. It is connected
with the fact that in all above considered cases the signs of
$S_\theta$ and $d\varphi/ds$ are opposite (the same relation holds
for the circular orbit with $r=3M$ \cite{40}). Naturally, when the
signs of $S_\theta$ and $d\varphi/ds$ coincide the spin-gravity
interaction acts as a attractive force (this type of
ultrarelativistic motions we shall consider in another
publication).


It is important that both the strict Mathisson-Papapetrou equations at the
Mathisson-Pirani supplementary condition and the shortened variant of these equations
admit the partial solutions in the form of the ultrarelativistic non-equatorial circular
orbits with constant latitude in a Schwarzschild field (however, the region of existence
of these orbits in the second case is much smaller, see equations (\ref{42}),
(\ref{56})). Moreover, on these solutions the expression $|S^{\mu\nu}P_\nu|$ in the
Tulczyjew-Dixon condition is much less than 1 due to the small value $\varepsilon$ which
is determined by (\ref{8}).

Naturally, the ultrarelativistic circular orbits of a spinning particle in a
Schwarzschild field are unstable. The dynamics of the deviation of a spinning particle
from the circular ultrarelativistic orbit with $r=3M$ in a Schwarzschild field caused by
the non-zero initial value of the radial velocity is investigated in section 5.

The result of importance is that spin can influence the shape of a trajectory very
strongly, as compare to geodesic trajectory, for the short time of the particle's motion.
Namely, for less than one or two revolutions of a spinning particle around a
Schwarzschild mass the difference of the radial coordinate $\Delta r$ of the spinning and
spinless particle becomes comparable with the initial radial coordinate of these
particles (see, for example, figures 1 and 6, or 4 and 8). We stress that this effect of
the considerable space separation of particles with different spin takes place just for
the small value $\varepsilon$ at the ultrarelativistic velocity with the Lorentz
$\gamma$-factor of order $\varepsilon^{-1/2}$.

In \cite{14} we read: "The simple act of endowing a black hole
with angular momentum has led to an unexpected richness of
possible physical phenomena. It seems appropriate to ask whether
endowing the test body with intrinsic spin might not also lead to
surprises". Now we can answer this question in the positive sense,
at least in the theoretical plane.

Perhaps, the further analysis of the gravitational ultrarelativistic spin-orbit
interaction, as well as of the gravitational ultrarelativistic spin-spin interaction (for
example, in a Kerr spacetime), will be useful for more fine investigations of the
gravitational collapse and cosmology problems.

\end{document}